\documentclass[twocolumn,aps,amsmath,emssymb,nofootinbib]{revtex4-1}
\usepackage{graphicx,txfonts}
\usepackage{color}
\begin{document}

\title{A self-consistent phase-space distribution function for the anisotropic Dark Matter halo of the Milky Way}

\author{Mattia Fornasa}
\affiliation{School of Physics and Astronomy, University of Nottingham, University Park, Nottingham, NG7 2RD, United Kingdom}
\author{Anne M. Green}
\affiliation{School of Physics and Astronomy, University of Nottingham, University Park, Nottingham, NG7 2RD, United Kingdom}

\begin{abstract}
Dark Matter (DM) direct detection experiments usually assume the simplest 
possible `Standard Halo Model' for the Milky Way (MW) halo in which the 
velocity distribution is Maxwellian. This model assumes that the MW halo is an 
isotropic, isothermal sphere, hypotheses that are unlikely to be valid in 
reality. An alternative approach is to derive a {\it self-consistent} solution 
for a particular mass model of the MW (i.e.~obtained from its gravitational 
potential) using the Eddington formalism, which assumes isotropy. In this 
paper we extend this approach to incorporate an anisotropic phase-space 
distribution function. We perform Bayesian scans over the parameters defining 
the mass model of the MW and parameterising the phase-space density, 
implementing constraints from a wide range of astronomical observations. 
The scans allow us to estimate the precision reached in the reconstruction of 
the velocity distribution (for different DM halo profiles). As expected, 
allowing for an anisotropic velocity tensor increases the uncertainty in the
reconstruction of $f(\mathbf{v})$, but the distribution can still be
determined with a precision of a factor of 4-5. The mean velocity 
distribution resembles the isotropic case, however the amplitude of the 
high-velocity tail is up to a factor of 2 larger. Our results agree with
the phenomenological parametrization proposed in Mao et al. (2013) as a good 
fit to N-body simulations (with or without baryons), since their velocity 
distribution is contained in our 68\% credible interval.
\end{abstract}

\maketitle

\section{Introduction}
\label{sec:Introduction}
The goal of Milky Way (MW) mass modelling is to build a model of our Galaxy 
in terms of the density distributions of its components \citep{Schmidt:1956,
Bahcall:1980fb,Caldwell:1981rj,Rohlfs:1988,Dehnen:1996fa,Olling:2000}. Mass 
models are the first step towards more complete dynamical descriptions of the 
MW in which the phase-space distribution $F(\mathbf{x},\mathbf{v})$ 
\citep{Binney_Tremaine}, consistent with the potential, is determined. The 
topic has recently received renewed interest (see, e.g., Refs. 
\citep{Klypin:2001xu,Widrow:2008yg,Sofue:2008wt,Catena:2009mf,Salucci:2010qr,
Catena:2011kv,Iocco:2011jz,McMillan:2011wd,Sofue:2011kw,Deg:2012eu,
Nesti:2013uwa,Burch:2013pda,Bozorgnia:2013pua}). This is in part due to the 
importance of accurate determinations of the local Dark Matter (DM) density, 
$\rho_{0}$, and circular velocity, $\Theta_{0}$, for Weakly Interacting Massive 
Particle (WIMP) direct detection experiments, which aim to detect the recoil 
energy deposited in a detector when WIMPs scatter off nuclei (see 
Refs. \citep{Cerdeno:2010jj,Strigari:2012gn,Peter:2013aha} for recent reviews).

The direct detection energy spectrum and its annual modulation, due to the 
Earth's orbit \citep{Drukier:1986tm}, depend on the local velocity 
distribution 
$f({\bf v}) = F(\mathbf{x_{\odot}},\mathbf{v})/\rho(\mathbf{x_\odot})$, where 
$\mathbf{x_\odot}$ denotes the position of the Sun and $\rho(\mathbf{x})$ is
the DM density (see Refs. \citep{Fairbairn:2008gz,Kuhlen:2009vh,Green:2010gw,
McCabe:2010zh,Green:2011bv,Fairbairn:2012zs}, among others). 
Direct detection experimental data are usually analysed assuming the 
so-called Standard Halo Model (SHM), which describes the MW as an isotropic 
isothermal sphere with local density 
$\rho_{0} = 0.3 \, {\rm GeV \, {\rm cm}}^{-3}$ and a Maxwellian-Boltzmann 
velocity distribution
\begin{equation}
\label{shm}
f({\bf v}) = \frac{1}{(2 \pi)^{3/2} \sigma^3} \exp{\left(- \frac{v^2}{2 \sigma^2} \right)} \,,
\end{equation}
with velocity dispersion $\sigma = \Theta_{0}/\sqrt{2}$ and 
$\Theta_{0}= 220 \, {\rm km \, s}^{-1}$. This model is unlikely to be a good 
approximation to the real MW DM halo. $N$-body simulations produce 
halos with velocity distributions which deviate systematically from a 
Maxwellian \citep{Vogelsberger:2008qb,Kuhlen:2009vh}.

Finding an appropriate phase-space distribution for the DM halo of the MW 
when you know its gravitational potential (i.e. obtaining a complete dynamical 
model for the Galaxy from its mass model) can be done under certain 
simplifying assumptions. For instance the phase-space distribution of a 
spherically symmetric system, with an isotropic velocity tensor, can be written
as function of the energy $E$ alone \citep{Binney_Tremaine}. In this case one 
can solve for $F(E)$, using the Eddington equation \citep{Eddington:1915}. 
The solution will be self-consistent, in that $F(E)$ and the gravitational
potential of the system, $\Phi(\mathbf{x})$, satisfies Boltzmann's equation 
(e.g. Refs. \citep{Catena:2011kv,Bhattacharjee:2012xm,Pato:2012fw}).

For the more general case of a spherical system with an anisotropic velocity 
tensor, the phase-space distribution function also depends on the modulus of
the angular momentum, $L$. Often a parametric form is considered for $F(E,L)$
and one can still find the set of parameters that corresponds to a 
self-consistent solution. Ref. \citep{Wojtak:2008mg} assumed that the 
phase-space distribution is separable in the two variables (i.e. 
$F(E,L)=F_E(E)F_L(L)$) and proposed a particularly convenient expression for 
$F_L(L)$ that depends on three parameters and can fit the radial dependence 
of the velocity anisotropy parameter $\beta(r)$
\begin{equation}
\beta(r) = 1 - \frac{\sigma_{\rm t}^2}{2\sigma_{\rm r}^2} \,,
\end{equation}
in the case of halos formed in $N$-body simulations, where $\sigma_{\rm t} $ 
and $\sigma_{\rm r}$ are the tangential and radial velocity dispersions.
In this paper, we apply the formalism developed in Ref. \citep{Wojtak:2008mg} 
in the context of the DM halos of galaxy clusters to the MW DM halo (see also 
recent work in Ref. \citep{Bozorgnia:2013pua} for an alternative approach to 
anisotropy). The first step is to build a mass model of the MW, c.f., e.g., 
Refs. \citep{Catena:2009mf,McMillan:2011wd,Catena:2011kv,Nesti:2013uwa}, using 
a wide range of astronomical observations to constrain the gravitational 
potential of the MW and, therefore, the DM density profile. Our inferred 
knowledge of $\Phi(\mathbf{x})$ will then be used to derive self-consistent 
solutions for $F(E,L)$, using the three-parameter form of $F_L(L)$ introduced 
in Ref. \citep{Wojtak:2008mg}.

This approach can be thought of as a generalization of the Eddington equation 
to the case of a system with an anisotropic velocity tensor. Moreover, it 
extends the approach used when constructing a mass model of the MW, where the
density profiles of the different components of the MW are written as 
functions of a number of free parameters which are constrained using 
astronomical observations. In this case, we also parametrize $F_L(L)$ and use 
our knowledge of the gravitational potential to derive self-consistent
solutions for $F(E,L)$, and therefore $f(\mathbf{v})$.

This is a different approach to that which has previously been used for 
anisotropic halos (e.g. Refs. \citep{Fairbairn:2008gz,Strigari:2009zb,
Fairbairn:2012zs}) where the components of the DM velocity dispersion tensor 
have been found by solving the Jeans equation \citep{Binney_Tremaine}. 
The velocity distribution is reconstructed with a remarkable precision, but 
the resulting solutions are not necessary self-consistent. In our approach, 
the Jeans equation is automatically satisfied (thanks to the Jeans theorem) 
without having to impose it explicitly.

The paper is structured as follows. In Sec. \ref{sec:MW} we introduce our mass 
model for the MW, listing the free parameters of the model in 
Sec. \ref{sec:mass_model} and the observations we use to constrain the 
parameters in Sec. \ref{sec:constraints}, while Sec. \ref{sec:sampling} 
describes the statistical techniques employed in the scans. 
Sec. \ref{sec:results} is devoted to the discussion of the resulting 
constraints on the mass model parameters. In 
Sec. \ref{sec:velocity_distribution} we present our technique for obtaining 
self-consistent anisotropic phase-space distributions, $F(E,L)$, and we apply 
it to the MW model found in the previous sections. In Sec. \ref{sec:discussion}
we discuss our results and in Sec. \ref{sec:conclusions} we summarize our 
conclusions.

\section{Mass models of the Milky Way}
\label{sec:MW}
In this section we discuss how we obtain a viable mass model for the MW. Our 
general approach follows previous work e.g., Refs. \citep{Catena:2009mf,
McMillan:2011wd,Catena:2011kv,Nesti:2013uwa}, with some differences in the 
details of how observations are implemented and in the modelling of the mass 
components.

The basic idea is to model the dynamically important components of the MW with 
physically motivated parametrisations, and then constrain the free parameters 
using a range of observations. We use a nested sampling algorithm to search 
the parameter space, and find the Bayesian probability distribution functions 
for the free parameters. If the observational data used are informative enough, 
the final results will be a precise MW model in agreement with observations, 
as well as estimates of the residual uncertainties in the model parameters.

In the follow subsection (Sec. \ref{sec:mass_model}) we describe the components
of our MW mass model, including the free parameters. In 
Sec. \ref{sec:constraints} we list and discuss the different 
observations and their implementation. Finally in Sec. \ref{sec:sampling} we 
give some details about the sampling technique.

\subsection{Milky-Way mass contributors}
\label{sec:mass_model}
Our mass model of the MW follows closely that proposed in Ref. 
\citep{Catena:2009mf}, and has four components: 
\begin{itemize}
\item {\bf stellar disk}: 
Following Ref. \citep{Catena:2009mf}, we model the stellar disk with a single 
component thin disc with density profile (in cylindrical coordinates):
\begin{equation}
\rho_{\rm d}(R,z) = \frac{\Sigma_{\rm d}}{2z_{\rm d}} \exp(-R/R_{\rm d}) \,
\mbox{sech}^2 \left( \frac{z}{z_{\rm d}} \right) \,,
\label{eqn:disk}
\end{equation}
which is in agreement with the fit to the COBE Diffuse Infrared Background 
Experiment data discussed in Ref. \citep{Freudenreich:1997bx}. The scale 
length in the $z$-direction is fixed at $z_{\rm d} = 0.34 \, {\rm kpc}$, since 
the dynamical constraints considered here are insensitive to small variations 
in its value, while the normalization, $\Sigma_{\rm d}$, and the radial scale 
length, $R_{\rm d}$, are left as free parameters. 

Ref. \citep{McMillan:2011wd} considered a mass model with two disks, a thin 
and a thick one. Since the stellar components are not the focus of our 
investigation, we consider a model with a single disk which has fewer free 
parameters (see also Sec. \ref{sec:local_surface_density}).

The gravitational potential produced by Eq.~(\ref{eqn:disk}) is axisymmetric 
(see, e.g. Ref. \citep{Burch:2013pda}), however, for simplicity, we work 
under the assumption of spherical symmetry, leaving the investigation of 
non-spherical Galactic models to future work. Thus, the disk gravitational 
potential at a certain distance $r$ from the center of the MW can be well 
approximated by $G\int_r^{\infty} {\rm d} r \, M_{\rm d}(<r)/r$, where 
$M_{\rm d}(<r)$ is the disk mass enclosed in a sphere of radius $r$. The 
deviation of the spherical gravitational potential from the true axisymmetric 
one (for the best-fit point for a Navarro-Frenk-White DM halo, see later) is 
maximal near the Galactic Center and is less than 10\% for distances larger 
than 2.2 kpc\footnote{This is the deviation with respect to the average of the 
axisymmetric and spherical rotation curves, where we calculate the 
axisymmetric case through a decomposition in spherical harmonics up to 
$\ell \leq 6$, see Ref. \citep{Binney_Tremaine}.}. 

\item {\bf bulge/bar}:
As in Ref. \citep{Catena:2009mf}, we consider an axisymmetric version of the 
model proposed in Ref. \cite{Zhao:1995qh}:
\begin{eqnarray}
\rho_{\rm bb}(x,y,z) &=& \rho_{\rm bb}(0)  \left[
s_{\rm a}^{-1,85} \exp(-s_{\rm a})  \right. \nonumber \\
& & \left. + \exp(-0.5 s_{\rm b}^2) \right] \,,
\label{eqn:bulge_bar}
\end{eqnarray}
with
\begin{equation}
s_{\rm a}^2 = \frac{q_{\rm b}^2(x^2 + y^2) + z^2}{z_{\rm b}^2} \,,
\end{equation}
and
\begin{equation}
s_{\rm b}^4 = \left( \frac{x^2 + y^2}{x_{\rm b}^2} \right)^2 + 
\left( \frac{z}{z_{\rm b}} \right)^4 \,.
\end{equation}
The two terms represent the bar and the bulge, respectively. Their parameters
are set to $q_{\rm b}=0.6$, $z_{\rm b}=0.4 \mbox{ kpc} \, (8 \mbox{ kpc} /R_0)$ 
and $x_{\rm b}=0.9 \mbox{ kpc} \, (8 \mbox{ kpc} /R_0)$, rescaling to arbitrary 
$R_0$, the distance of the Sun from the Galactic Center, as suggested by 
Ref. \citep{Zhao:1995qh}. The normalization $\rho_{\rm bb}(0)$ is left free. 
Another viable model for the bulge/bar system can be found in 
Ref. \citep{Bissantz:2001wx}. This component is always subdominant in our 
results and therefore we do not expect our results for the local DM 
distribution to be sensitive to the details of the bulge/bar density 
parametrization. As for the disk, the gravitational potential of the bulge/bar 
is assumed to be spherical and is obtained by computing the enclosed mass.

\item {\bf interstellar medium}: 
The model for the interstellar medium is kept fixed, without any free 
parameters. The mass density of molecular hydrogen $H_2$, as well as the
HI and HII components, is modelled as in Ref. \citep{Moskalenko:2001ya}, based 
on the observations presented in Ref. \cite{Gordon:1976}.

\item {\bf DM halo}:
Insight into the density profile of the DM halo comes mainly from the study of 
synthetic halos formed in $N$-body simulations \citep{Springel:2008cc,
Navarro:2008kc,Diemand:2008in,Diemand:2009bm}. We consider three different 
spherically symmetric parametrizations for the DM density profile: a 
Navarro-Frenk-White (NFW) profile \citep{Navarro:1996gj}  
\begin{equation}
\rho_{\chi}(r) = \rho_{\rm s} \left( \frac{r}{r_{\rm s}} \right)^{-1} 
\left( 1+\frac{r}{r_{\rm s}} \right)^{-2},
\label{eqn:NFW}
\end{equation}
an Einasto profile
\citep{Einasto:1965}
\begin{equation}
\rho_{\chi}(r) = \rho_{\rm s} \exp \left( - \frac{2}{\alpha} 
\left[ \left( \frac{r}{r_{\rm s}} \right)^\alpha - 1 \right] \right),
\label{eqn:Einasto}
\end{equation}
and a Burkert profile \citep{Burkert:1995yz} 
\begin{equation}
\rho_{\chi}(r)=\rho_{\rm s} \left( 1 + \frac{r}{r_{\rm s}} \right)^{-1}
\left( 1 + \frac{r^2}{r_{\rm s}^2} \right)^{-1}.
\label{eqn:Burkert}
\end{equation}
The scale radius, $r_{\rm s}$, is related to the radius at which the 
logarithmic derivative of the density profile is equal to $-2$, while
$\rho_{\rm s}$ fixes the normalization. 
The NFW and Burkert profiles only have two free parameters ($r_{\rm s}$ and 
$\rho_{\rm s}$) while the Einasto profile has an additional parameter, 
$\alpha$, which controls the curvature of the profile. The NFW and Einasto 
profiles have inner cusps and provide good fits to the density profiles of 
halos formed in DM-only $N$-body simulations. Baryons are likely to play an 
important role in determining the DM distribution in the inner regions. 
However, simulating baryonic physics, and forming realistic galaxies, is a 
difficult problem (see, e.g., Refs. \citep{Guedes:2011ux,Marinacci:2013mha} for 
recent progress) and it is not yet clear how baryonic physics will affect the 
DM density profile. The Burkert profile has a central core, a possibility that 
seems to be preferred by observations of dwarf Spheroidal \citep{Walker:2011zu}
or Low-Surface Brightness galaxies \citep{deBlok:2002tg}.

To compare with other mass models and other DM halo constraints present in the 
literature, we will also calculate the concentration parameter 
$c = r_{\rm vir}/r_{\rm s}$, where $r_{\rm vir}$ is the virial radius, i.e. the 
radius within which the average density of the halo is $\Delta$, the virial 
overdensity, times the critical density of the Universe. In a flat 
$\Lambda$ Cold DM Universe with matter density parameter $\Omega_{\rm m}=0.227$ 
\citep{Jarosik:2010iu} at $z=0$, $\Delta=94$ 
\citep{Bryan:1997dn}\footnote{Using the more recent value of $\Omega_{\rm m}$ 
from the Planck collaboration would not affect our results significantly.}.
\end{itemize}

Unlike Ref. \citep{Catena:2011kv} we do not include uncollapsed baryons in our
MW model, as their distribution is highly uncertain and the majority are 
thought to be in the warm-hot intergalactic medium (see e.g. Ref.
\citep{Bregman:2009py}). See Sec. \ref{sec:results} for further discussion.

There are five additional quantities that will be needed when implementing the 
observational constraints (see Sec. \ref{sec:constraints}). These are $R_0$, 
$\beta_\ast$ (the velocity anisotropy of stellar halo tracers, see 
Sec. \ref{sec:velocity_dispersion}) and the three components of the velocity 
of the Sun with respect to the Rotation Standard of Rest (see 
Sec. \ref{sec:circular_velocity}), $\mathbf{V_\odot^{\mbox{\tiny{RSR}}}}=
(U_\odot^{\mbox{\tiny{RSR}}}, V_\odot^{\mbox{\tiny{RSR}}}, W_\odot^{\mbox{\tiny{RSR}}})$, in a 
system of coordinates where the first axis points towards the Galactic Center, 
the second along the direction of Galactic rotation and the third is
perpedicular to the Galactic plane. 

\subsection{Experimental constraints}
\label{sec:constraints}
As emphasised by Ref. \citep{McMillan:2011wd} the fact that different 
experimental constraints use different underlying assumptions is an issue when 
constructing a MW mass model. In principle the best approach would be to use 
the raw data, rather than values of derived quantities, however in practice 
this is not possible. Still, where possible, we avoid using constraints which 
make specific assumptions, e.g. a fixed value of the solar radius, $R_{0}$.

\subsubsection{Local circular velocity}
\label{sec:circular_velocity}
Measurements of the local circular velocity, $\Theta_0$, can be divided
into two categories: those that measure the rotation velocity of the Sun, 
$V_{\phi,\odot}$, by observing the proper motion of an object (or a population of 
objects) at rest at the Galactic Center, and those that measure the difference 
between the two Oort constant, $A-B=\Theta_0/R_0$, from the proper motions of 
tracers.

Ref. \cite{Reid:2004rd} measured the proper motion of Sgr A$^\star$ with an 
extremely good accuracy of approximately 0.4\%: $\mu_l = (-6.379 \pm 0.026) \,  
{\rm mas \, yr}^{-1}$. The local circular velocity can then be calculated 
using $R_{0}$, and the velocity of the Sun with respect to the so-called 
Rotation Standard of Rest (RSR), $V_{\odot}^{\mbox{\tiny{RSR}}}$, i.e. the rotation 
velocity of a circular orbit in the axisymmetric approximation of the 
gravitation potential \citep{Shuter:1982,Bovy:2012ba}. On the other hand, 
Ref. \citep{Feast:1997sb} measured $A-B$ with 3\% accuracy using the motion 
of 220 Cepheids detected by the Hipparcos satellite: 
$A-B = (27.2 \pm 0.9) \, {\rm km \, s}^{-1} {\rm kpc}^{-1}$.

The two techniques appear to lead to inconsistent values of $\Theta_0$, 
depending on the value of $V_\odot^{\mbox{\tiny{RSR}}}$ assumed. Traditionally it 
has been assumed that the Local Standard of Rest (LSR, i.e. the orbit of local 
stars with ``zero velocity dispersion'', obtained by extrapolating to 
$\sigma_{\rm R}=0$ the definition of the asymmetric drift, 
\citep{Fich:1991ej,Bovy:2012ba}) moves on a circular orbit. If this is true 
then the rotational component of the Sun's velocity with respect to the LSR, 
$V_{\odot}^{\mbox{\tiny{LSR}}}$, coincides with $V_{\odot}^{\mbox{\tiny{RSR}}}$, and can be 
used to estimate $\Theta_0$ from the measurement of $V_{\phi,\odot}$. Using 
$V_{\odot}^{\mbox{\tiny{LSR}}} = 5 \, {\rm km \, s}^{-1}$, from the analysis in 
Ref. \citep{Dehnen:1996fa} of Hipparcos data and $R_0=8.0$ kpc from 
Ref. \citep{Reid:1993fx}, Ref. \cite{Reid:2004rd} find 
$\Theta_0/R_0 = (29.4 \pm 0.2) \, {\rm km \, s}^{-1} {\rm kpc}^{-1}$, 
significantly larger than the value quoted in Ref. \citep{Feast:1997sb} of 
$(27.2 \pm 0.9) \, {\rm km \, s}^{-1} {\rm kpc}^{-1}$.
However, using line-of-sight velocities of more than 3000 stars observed by 
the APOGEE survey, Ref. \citep{Bovy:2012ba} found 
$V_{\odot}^{\mbox{\tiny{RSR}}} = V_{\phi,\odot}-\Theta_0 = 23.9^{+5.1}_{-0.5} \, {\rm km \, s}^{-1}$ 
(assuming a flat rotation curve). This is significantly larger than even the 
revised value of $V_{\odot}^{\mbox{\tiny{LSR}}}$ of $13 \, {\rm km \, s}^{-1}$ 
\citep{Schoenrich:2010}, found taking into account the radial metallicity 
gradient. Ref. \citep{Bovy:2012ba} discussed two possible reasons for this 
discrepancy. For instance, the LSR would differ from the RSR if the orbit of 
the LSR is not circular (due, for instance, to large-scale non-axisymmetric 
streaming motions). Alternatively, $V_{\odot}^{\mbox{\tiny{LSR}}}$ could be 
significantly larger than previously thought.

Using the value of $V_{\odot}^{\mbox{\tiny{RSR}}} = 23.9 \, {\rm km \, s}^{-1}$ quoted 
in Ref. \citep{Bovy:2012ba}, the value of $\Theta_0/R_0$ found from the 
measurement of the proper motion of Sgr A$^\star$ in Ref. \cite{Reid:2004rd} 
drops to $27.1 \, {\rm km \, s}^{-1} {\rm kpc}^{-1}$, consistent with both the 
value in Ref. \citep{Feast:1997sb} value and the measurement of 
$\Theta_{0}$ from Ref. \citep{Bovy:2012ba}.

In light of this, we constrain the circular velocity by imposing the
measurement of the proper motion of Sgr A$^\star$ by Ref. \cite{Reid:2004rd},
assuming the value of $V_{\odot}^{\mbox{\tiny{RSR}}}$ quoted before from 
Ref. \cite{Bovy:2012ba} (see also Sec. \ref{sec:nuisance}). We also follow 
Ref. \citep{Catena:2009mf} by using 
$A+B = \partial \Theta(r) / \partial r|_{r=R_0} = (0.18 \pm 0.47) \, 
{\rm km \, s}^{-1} {\rm kpc}^{-1}$.

\subsubsection{Local surface density}
\label{sec:local_surface_density}
The total integrated local surface density, within a vertical distance $z$ of 
the Galactic plane, is defined as
\begin{equation}
\Sigma(R_0,z)= \int_{-z}^z \rho_{\mbox{\tiny{tot}}}(R_0,z) \, {\rm d} z \,,
\end{equation}
where $\rho_{\mbox{\tiny{tot}}}$ is the total mass density of the MW.
A demonstration that this quantity continues increasing for $z$ larger 
than a few times the disk scale length, $z_{\rm d}$, would be very strong 
evidence for the presence of DM at the Solar radius.
The local surface density can be estimated by means of the Poisson equation, 
once the vertical force, $K_z$, is determined. We use the values, derived from 
the kinematics of stellar tracers, from Ref. \citep{Kuijken:1989} and 
Ref. \citep{Kuijken:1991} of 
$\Sigma_{\ast}(R_0)=(48 \pm 8) M_\odot \mbox{pc}^{-2}$ and 
$\Sigma(R_0,1.1 \mbox{ kpc})=(71 \pm 6) M_\odot \mbox{pc}^{-2}$ for the visible
component and the total mass within 1.1 kpc, respectively.
These values are consistent with more recent analyses, e.g. Refs.
\citep{Holmberg:1998xu,Siebert:2002nx,Holmberg:2004fj,Bovy:2011zx,
Zhang:2012rsb}.

Ref. \citep{Bovy:2012tw} used the data from 
Refs. \citep{Bidin:2010rj,Bidin:2012za,Bidin:2012vt} to derive lower limits on 
the local surface density up to 4 kpc. We do not use these results since they 
are, strictly speaking, only lower limits and also because of the large
residuals in the fit to $\overline{UW}$ (see Fig. 2 of 
Ref. \citep{Bidin:2010rj}).

\subsubsection{Terminal velocities}
The inner rotation curve of the MW, i.e. inside the Solar radius, can be 
constrained by measurements of the so-called terminal velocities: along each 
line-of-sight towards Galactic longitude $l$ there is a point at which the 
distance from the Galactic Centre is smallest.
Under the assumption of circular motion, the modulus of the line-of-sight 
velocity is largest for objects at this minimum distance that are moving on an 
orbit that is tangential to the line-of-sight. This maximal velocity is 
normally referred to as the {\it terminal velocity} and can be used to 
directly constrain the rotation curve of the MW, at that specific minimal 
distance.

Measurements of terminal velocities are obtained from the observation of
the spectral line of atomic hydrogen HI (see, e.g. 
Ref. \citep{Malhotra:1994qj}) or of CO \citep{McClureGriffiths:2007ts}. We 
consider the data set in Ref. \citep{Malhotra:1994qj}, excluding all the 
points with $|\sin l|<0.35^\circ$, where the assumption of circular motion is 
not valid due to the presence of the Galactic bar. A constant experimental 
error of 7 ${\rm km \, s}^{-1}$ is assumed for each of the remaining data 
points (following Ref. \citep{Catena:2009mf}).

\subsubsection{Microlensing}
Microlensing observations constrain the gravitational potential of the MW 
since they provide us with a probe of the mass density in compact objects in 
the direction(s) of observation.
The impact of microlensing data on the reconstruction of the MW potential has 
been studied in Ref. \citep{Iocco:2011jz}. We consider the same 10 
measurements of the optical depth $\langle \tau \rangle$ discussed in that 
paper, coming from the MACHO \citep{Popowski:2004uv}, OGLE-II 
\citep{Sumi:2005da} and EROS \citep{Rahal:2009yt} collaborations. The 
distribution of the gravitational lenses is assumed to follow the matter 
density of the disk and the bulge/bar, while the distribution of sources 
depends on the particular microlensing events (see Ref. \citep{Iocco:2011jz} 
for more details).

\subsubsection{Proper motion of masers in high-mass star-forming regions}
Accurate measurements of the proper motion ($\mu_{\rm l}$ and $\mu_{\rm b}$) and 
of the line-of-sight velocity with respect to the LSR ($v_{\mbox{\tiny{LSR}}}$) 
are available in the literature for a number of masers. Their positions can
also be determined from their Celestial coordinates (which we consider
to be exact) and their parallaxes, $\pi$.
The masers are found to be in almost circular orbits and 
Ref. \citep{Reid:2009nj} showed how, once one fixes the peculiar (i.e. 
non-circular) velocity $(U_{\rm s},V_{\rm s},W_{\rm s})$ of the maser, it is 
possible to predict its proper motion. The transformation from peculiar to 
proper motion requires knowledge of the position of the maser, the 
circular velocity at the position of the maser and the velocity of the Sun 
with respect to the RSR. 
The latter is included in our nuisance parameters (see Sec. \ref{sec:nuisance}) 
so that, for each choice of $(U_{\rm s},V_{\rm s},W_{\rm s})$, the predicted proper 
motion can be compared to the measured values for $\mu_{\rm l}$, $\mu_{\rm b}$ and 
$v_{\mbox{\tiny{LSR}}}$, providing an indirect constraint on the rotation curve at 
the position of the maser.

Ref. \citep{Reid:2009nj} found that the 18 masers they analyzed were orbiting  
around the Galactic Center with (on average) 
$V_{\rm s} \sim -15 \, {\rm km \, s}^{-1}$. However Ref. \citep{McMillan:2009yr} 
argue that it is more likely that
$V_{\odot}^{\mbox{\tiny{LSR}}} \approx 11 \, {\rm km \, s}^{-1}$, as advocated by 
Ref. \citep{Schoenrich:2010}, rather than the value of 
$V_{\odot}^{\mbox{\tiny{LSR}}} \approx 5 \, {\rm  km \, s}^{-1}$ \citep{Dehnen:1996fa}
used by Ref. \citep{Reid:2009nj}. Note that the value of 
$V_{\odot}^{\mbox{\tiny{LSR}}}$ proposed in Ref. \citep{Schoenrich:2010} is still 
smaller than the value in Ref. \citep{Bovy:2012ba} of 
$V_{\odot}^{\mbox{\tiny{RSR}}} = 23.9 \, {\rm km \, s}^{-1}$ which we adopt, see 
Sec. \ref{sec:circular_velocity}.

We implement the information from the motions of masers, following 
Ref. \citep{McMillan:2009yr}, by marginalizing over the peculiar velocities
and parallaxes of the masers. We assume that the components of the peculiar 
velocities have a Gaussian probability distribution with zero mean and 
$\Delta v=10 \, {\rm km \, s}^{-1}$. A Gaussian distribution is also assumed 
for the parallax, with the mean and dispersion coinciding with the measured 
value of $\pi$ and  the experimental error, respectively, for each maser.

We consider a total of 33 masers from Refs. \citep{Reid:2009nj,Rygl:2009rc,
Hachisuka:2009hz,Sanna:2009tt,Sato:2010xh,Sato:2010dr,Ando:2010by,
Sanna:2011ie,Nagayama:2011fd,Xu:2011fk,Sakai:2012qs}.

\subsubsection{Velocity dispersion of halo stars}
\label{sec:velocity_dispersion}
Ref. \citep{Xue:2008se} selected a sample of more than 2000 Blue 
Horizontal-Branch stars observed by the Sloan Digital Sky Survey and computed 
the dispersion of the line-of-sight velocity in 10 bins in distance from the 
Galactic Center, from 5.0 to 60.0 kpc. They compare these measurements with 
the results of two cosmological galaxy formation simulations of MW-like 
galaxies to infer the rotation curve of the MW up to 60.0 kpc.

We follow Ref. \citep{Catena:2009mf} and directly use the binned line-of-sight
velocity dispersion data. We do not use the results of the simulations,
however our analysis is based on the following three assumptions, that receive 
validation from the simulations used in Ref. \citep{Xue:2008se}: $i)$ the 
dispersion of the line-of-sight velocity can be used as an estimate of the 
radial velocity dispersion, $ii)$ the stellar density $\rho_\ast$ is well 
fitted by a $r^{-3.5}$ power law and $iii)$ the stellar velocity anisotropy 
parameter $\beta_\ast$ is constant over the range of distances considered. 
Under these assumptions, one can solve the Jeans equations for the Blue 
Horizontal-Branch stars analyzed by Ref. \citep{Xue:2008se} to obtain an 
expression for the radial velocity diversion of the stars \citep{Catena:2009mf}:
\begin{equation}
\sigma_{{\rm r},\ast}^2(r) = \frac{1}{r^{2\beta_{\ast}} \rho_\ast(r)} 
\int_r^\infty {\rm d}s \, s^{2\beta_\ast-1} \rho_\ast(s) \,
\Theta^2(s) \,.
\end{equation}
The 10 velocity dispersion data points in Ref. \citep{Xue:2008se} can thus be 
used to constrain the circular velocity at large radii. Note that the constant 
velocity anisotropy parameter of the stars, $\beta_\ast$, is a free parameter 
in our scans.

\subsubsection{Nuisance parameters}
\label{sec:nuisance}
The Sun's distance from the Galactic Center, $R_0$, as well as the three 
components of the velocity of the Sun with respect to the RSR are included in 
the scan as nuisance parameters. We assume the $U$ and $W$ components of the 
Sun's velocity with respect to the RSR and LSR are identical (see 
Sec. \ref{sec:circular_velocity}). We also assume a Gaussian probability 
distribution for each nuisance parameter, with a mean value and dispersion 
corresponding to the measured value and experimental error, respectively: 
\begin{itemize}
\item $R_0=(8.33 \pm 0.35)$ kpc, inferred from the observation of stellar 
orbits around the Galactic Center \citep{Gillessen:2008qv}.
\item  $U_{\odot}^{\mbox{\tiny{RSR}}}=(11.1 \pm 1.0) \, {\rm km \, s}^{-1}$ 
\citep{Schoenrich:2010}.
\item  $V_{\odot}^{\mbox{\tiny{RSR}}}=(23.9 \pm 5.1) \,  {\rm km \, s}^{-1}$ 
\citep{Bovy:2012ba}, see Sec. \ref{sec:circular_velocity}.
\item $W_{\odot}^{\mbox{\tiny{RSR}}}=(7.25 \pm 0.50) \, {\rm km \, s}^{-1}$ 
\citep{Schoenrich:2010}.
\end{itemize}

\subsubsection{Other observations}
\label{sec:other}
Refs. \citep{Wilkinson:1999hf,Battaglia:2005rj} estimated the total 
mass of the MW from the kinematics of satellite galaxies, globular clusters 
and (in the case of Ref. \citep{Battaglia:2005rj}) individual stars, considered 
as tracers of the underlying gravitational potential. Their results have an 
accuracy of approximately a factor of 2, pointing towards a total DM halo mass 
of around $\sim 10^{12} M_\odot$. However these estimates are obtained using 
assumptions which we do not make in our analysis (e.g. a fixed value of $R_0$
and of the Solar RSR velocity). Moreover, the results in 
Ref. \citep{Wilkinson:1999hf} become very prior dependent as soon as the value 
of $\Theta_0$ is left free, as it is in our scans since it depends on the mass 
model. Thus, we decide not to consider such results.

Ref. \citep{Piffl:2013mla} found, using high velocity stars from the RAVE 
survey, that the local escape velocity lies between 492 and 587 
${\rm km \, s}^{-1}$ (at 90\% confidence). No constraint on the escape velocity 
is included in our scan since Ref. \citep{McMillan:2011wd} argued that 
assuming that these stars are in a steady state (as done in 
Ref. \citep{Piffl:2013mla}) is probably unrealistic. Moreover, the range 
quoted could be even larger depending on the parameterisation of the high 
speed tail of the stellar velocity distribution.

Finally, the mass model in Ref. \citep{McMillan:2011wd} used 
two additional constraints from simulations. Namely the 
concentration of MW-like halos from Ref. \citep{BoylanKolchin:2009an} and 
the following relation for the ratio of stellar mass, $M_{\star}$, to DM 
mass, $M_{\chi}$,:
\begin{equation}
\frac{M_\star}{M_{\chi}} = 0.129 \left[ 
\left( \frac{M_\chi}{10^{11.4}} \right)^{-0.926} + 
\left( \frac{M_\chi}{10^{11.4}} \right)^{0.129} \right]^{-2.44}
\label{eqn:ratio},
\end{equation} 
found by Ref. \citep{Guo:2009fn} to be a good fit to the Millennium-II 
simulation. We prefer to include only observational data and thus we do not 
consider the information coming from $N$-body simulations. However, in 
Sec. \ref{sec:results} we will show the predictions of our mass models for the 
observables discussed here (i.e. total mass of the MW within 50 kpc and 
within the virial radius, the escape velocity, the concentration of the MW DM 
halo and the DM/stellar mass ratio) but not included as data.

\subsection{Sampling technique}
\label{sec:sampling}
The free parameters that we scan over are summarised in 
Tab. \ref{tab:parameters}. The parameters of our MW mass model were introduced 
in Sec. \ref{sec:mass_model} (note that the quantity $\alpha$ is only 
applicable in the case of the Einasto profile). The distance of the Sun from 
the Galactic centre, $R_0$, and the components of the Sun's motion with 
respect to the RSR, $U_{\odot}^{\mbox{\tiny{RSR}}}$, $V_{\odot}^{\mbox{\tiny{RSR}}}$ and 
$W_{\odot}^{\mbox{\tiny{RSR}}}$, are considered as nuisance parameters (see 
Sec. \ref{sec:nuisance}). The last three parameters, $\beta_0$, $\beta_\infty$ 
and $k_{L_0}$, parameterise the part of the phase-space distribution function
which depends on the angular momentum and will be introduced in detail in
Sec. \ref{sec:velocity_distribution}. $\beta_0$ and $\beta_\infty$ are the 
velocity anisotropy parameters at the center of the MW and at infinity,
respectively. The third parameter entering in the $L$-dependent part of the
phase-space density is a characteristic scale $L_0$. The parameter considered
in the scan, however, is the ratio between $L_0$ and ``scale angular momentum''
$L_{\rm s} = r_{\rm s} \, V_{\rm s} = r_{\rm s} \, \Theta(r_{\rm s}) (\ln 2 - 1/2)^{-1/2}$.

The second column indicates the range of values considered for each of the 
parameters and the third column indicates the shape of the prior probability 
distribution.

\begin{table}
\begin{center}
\begin{tabular}{ccc}
{\bf Name} & {\bf Range} & {\bf Probability distribution} \\
\hline
$\rho_{\rm s}$ [$M_\odot \, {\rm pc}^{-3}$] & $10^{-5}$, 1.0 & log \\
$r_{\rm s}$ [${\rm kpc}$] & 0.0, 100.0& log \\
$R_{\rm d}$ [${\rm kpc}$] & 1.2, 4.0 & log \\
$\Sigma_{\rm d}$ [$M_\odot \, {\rm pc}^{-2}$] & $10^2$, $10^4$ & log \\
$\rho_{\rm bb}(0)$ [$M_\odot \, {\rm pc}^{-3}$] & $10^{-4}$, $10^2$ & log \\
$\alpha$ & 0.1, 0.4 & flat \\
\hline
$\beta_\ast$ & -1.0, 0.7 & flat \\
$R_0$ [${\rm kpc}$] & 6.5, 9.0 & Gaussian \\
$U_{\odot}^{\mbox{\tiny{RSR}}}$ [${\rm km \, s}^{-1}$] & 1.1, 21.1 & Gaussian \\ 
$V_{\odot}^{\mbox{\tiny{RSR}}}$ [${\rm km \, s}^{-1}$] & -7.76, 32.24 & Gaussian \\ 
$W_{\odot}^{\mbox{\tiny{RSR}}}$ [${\rm km \, s}^{-1}$] & 2.25, 12.25 & Gaussian \\
\hline
$\beta_0$ & -0.5, - & flat \\
$\beta_\infty $ & -0.5, 1.0 & flat \\
$k_{L_0}$ & $10^{-3}$, $10^3$ & log \\
\hline
\end{tabular}
\caption{\label{tab:parameters} Parameters defining the multidimensional space over which our scan is performed. The first column is the name of the parameter, the second the range of values considered in the scan and the third indicates the form of the prior probability distribution function. The first section of the table contains the parameters of the MW mass model introduced in Sec. \ref{sec:MW}:  $\rho_{\rm s}$ fixes the normalization of the DM halo and $r_{\rm s}$ is its scale radius. $R_{\rm d}$ is the radial scale density of the disc, $\Sigma_{\rm d}$ and $\rho_{\rm bb}(0)$ determine the normalization of the density profiles of the disk and bulge/bar respectively. Finally, $\alpha$ appears in the Einasto DM halo profile. The second section contains parameters which are needed to implement the astronomical constraints (see Sec. \ref{sec:constraints}): $\beta_\ast$ is the velocity anisotropy parameter for the Blue Horizontal-Branch stars considered in Ref. \citep{Xue:2008se} (see Sec. \ref{sec:velocity_dispersion}), $R_0$ is the distance of the Sun from the Galactic Center and ($U_{\odot}^{\mbox{\tiny{RSR}}}$, $V_{\odot}^{\mbox{\tiny{RSR}}}$, $W_{\odot}^{\mbox{\tiny{RSR}}}$) are the components of the velocity of the Sun with respect to the RSR. The final section contains the parameters which appear in the part of the phase-space distribution function which depends on the angular momentum: $\beta_0$ and $\beta_\infty$ are the velocity anisotropy of the DM halo at $r=0$ and $r=\infty$ respectively and $k_{L_0}$ is a proportionality constant between the parameter $L_0$ (entering in the definition of the phase-space density) and $r_s \, \Theta(r_s) \sqrt{\ln 2 - 1/2}$ (see Sec. \ref{sec:velocity_distribution}). We do not indicate here the upper limit for $\beta_0$ since it depends on the form of the density profile (see Sec. \ref{sec:velocity_distribution}).}
\end{center}
\end{table}

The scan is performed using the public code MultiNest \citep{Feroz:2013hea}, 
which uses a nested sampling algorithm to determine the Bayesian posterior 
probability distributions for the parameters in the scan and functions of 
these parameters. The core principle is Bayes' theorem: 
\begin{equation}
\mbox{Pr}(\mathbf{\Theta}|\mathbf{D},H) = 
\frac{\mbox{Pr}(\mathbf{D}|\mathbf{\Theta},H) \, \mbox{Pr}(\mathbf{\Theta}|H)}
{\mbox{Pr}(\mathbf{D}|H)},
\end{equation}
which shows how the so-called {\it prior probability} 
$\mbox{Pr}(\mathbf{\Theta}|H)$ of the parameters $\mathbf{\Theta}$ (describing 
our knowledge of the quantities in the context of model $H$ before the 
implementation of the experimental data $\mathbf{D}$) is updated once we 
consider the observational information encoded in the {\it likelihood} 
$\mathcal{L}(\mathbf{\Theta}) = \mbox{Pr}(\mathbf{D}|\mathbf{\Theta},H)$.
The result is the posterior probability distribution function (pdf)
$\mbox{Pr}(\mathbf{\Theta}|\mathbf{D},H)$, which also depends on the so-called
{\it evidence} $\mbox{Pr}(\mathbf{D}|H)$. We do not need to include the 
evidence as it only depends on the data and therefore acts as a normalization
constant.

Different choices of prior distributions can affect the final posterior pdf if
the data considered are not constraining enough to overcome the shape of the 
prior. The third column in Tab. \ref{tab:parameters} indicates for each
parameter whether we assumed a Gaussian prior distribution (denoted by 
{\it Gaussian}) or a uniform distribution on a linear scale ({\it flat}) or 
logarithmic scale ({\it log}). In the case of the NFW DM halo for some 
parameters we used both flat and log priors, in order to check that the 
posterior pdfs do not depend significantly on the choice of the prior. When
not explicitly mentioned, all the results presented in the following sections
refer to the choice of priors indicated in Tab. \ref{tab:parameters}.

We are interested in the probability distributions of particular subsets 
(normally 1- or 2-dimensional) of the parameters in Tab. \ref{tab:parameters}.
Two different statistics can be considered to measure such distributions. The 
so-called {\it posterior pdf of parameter $\Theta_i$} is found by 
marginalizing over all the other parameters:
\begin{eqnarray}
\mbox{pdf}(\Theta_i) & = & \int {\rm d}\Theta_0 \, {\rm d}\Theta_1 \cdots 
{\rm d}\Theta_{i-1} \, {\rm d}\Theta_{i+1} \cdots {\rm d}\Theta_{N-1} \, 
{\rm d}\Theta_{N}  \nonumber \\
&\times & \mbox{Pr}(\mathbf{\Theta}|\mathbf{D},H) \,,
\end{eqnarray}
where $N$ is the total number of parameters in the scan. On the other hand, the 
{\it profile likelihood (PL) of parameter $\Theta_i$} is found by maximizing
over the other parameters\footnote{One can write similar expressions for the 
pdf and PL of more than one parameter or for functions of $\Theta$.}:
\begin{equation}
\mbox{PL}(\Theta_i) = \max_{\Theta_0, \, \Theta_1, \cdots, \Theta_{i-1}, \, \Theta_{i+1}, \cdots, \Theta_{N-1}, \, \Theta_{N}} \mathcal{L}(\mathbf{\Theta}) \,.
\end{equation}
The PL is very sensitive to fine-tuned regions in the parameter space with
a very large likelihood, while the pdf takes into account volume effects where
large regions with a moderate likelihood are integrated over. It is normally
useful to consider both quantities when studying the characteristics of a
parameter space \citep{Feroz:2011bj,Strege:2012bt}.

For each parameter we will compute the so-called $x\%$ 
{\it credible interval}, defined so that the each of the two tails outside 
the interval has a probability $0.5x\%$. We also determine the $x\%$ 
{\it confidence levels} as the region with a likelihood at most 
$\Delta\chi^2(x)$ lower than the likelihood of the best-fit point, where 
$\Delta\chi^2(x)$ is determined by solving
\begin{equation}
x = \int_0^{\Delta\chi^2} {\rm d} y \, \chi^2_n(y) \,,
\end{equation}
where $\chi^2_n(y)$ is the $\chi^2$ distribution with $n$ degrees of 
freedom.  We used the SuperBayes 
package\footnote{http://www.ft.uam.es/personal/rruiz/superbayes/}
\citep{deAustri:2006pe,Trotta:2008bp} to compute the credible intervals and 
confidence regions and produce the plots presented in the following sections.

We performed our scans using 2000 live-points and a tolerance of $10^{-4}$.
Ref. \citep{Feroz:2011bj} showed that such a set-up allows a good 
reconstruction of the PL in the context of SuperSymmetric models of Particle 
Physics. The pdf and PL distributions for our scans do not differ much from 
each other, which confirms that we have achieved a reliable evaluation of the 
PL.

\section{Parameter estimation for the Milky Way mass models}
\label{sec:results}
In this section we present the results of our scans, determining the parameters 
of our MW mass model.
Tabs. \ref{tab:results_NFW}, \ref{tab:results_Einasto} and 
\ref{tab:results_Burkert} summarize the results, including mean and best-fit 
values and $68\%$ and $95\%$ credible intervals, for the NFW, Einasto and 
Burkert DM halo profiles, respectively. Note that in order to facilitate 
comparisons with other studies we include derived values of quantities 
(marked with $\ast$) that are not considered as experimental data in our scans. 

\begin{table*}
\begin{center}
\begin{tabular}{cccccccc}
\multicolumn{8}{c}{\bf NFW} \\
\hline
{\bf Name} & {\bf Measured value} & {\bf mean} & {\bf best-fit} & {\bf lower 68\%} & {\bf upper 68\%} & {\bf lower 95\%} & {\bf upper 95\%} \\
\hline
$\rho_{\rm s}$ [$M_\odot \, {\rm pc}^{-3}$] & & $10^{-1.86}$ & $10^{-1.53}$ & $10^{-2.21}$ & $10^{-1.51}$ & $10^{-2.56}$ & $10^{-1.30}$ \\
$r_{\rm s}$ [${\rm kpc}$] & & 17.31 & 10.01 & 9.89 & 24.43 & 7.79 & 41.64 \\
$R_{\rm d}$ [${\rm kpc}$] & & 2.44 & 2.56 & 2.23 & 2.66 & 2.07 & 2.94 \\
$\Sigma_{\rm d}$ [$M_\odot \, {\rm pc}^{-2}$] & & $10^{3.10}$ & $10^{3.03}$ & $10^{3.00}$ & $10^{3.20}$ & $10^{2.88}$ & $10^{3.26}$ \\
$\rho_{\rm bb}(0)$ [$M_\odot \, {\rm pc}^{-3}$] & & $10^{-2.15}$ & $10^{-2.79}$ & $10^{-3.38}$ & $10^{-0.87}$ & $10^{-3.89}$ & $10^{-0.15}$ \\
$\beta_\ast$ & & -0.44 & -0.23 & -0.74 & -0.15 & -0.94 & 0.08 \\
$R_0$ [${\rm kpc}$] & & 8.13 & 8.20 & 7.85 & 8.41 & 7.57 & 8.67 \\
$U_{\odot}^{\mbox{\tiny{RSR}}}$ [${\rm km \, s}^{-1}$] & & 11.08 & 10.95 & 10.17 & 12.00 & 9.27 & 12.89 \\ 
$V_{\odot}^{\mbox{\tiny{RSR}}}$ [${\rm km \,s}^{-1}$] & & 25.44 & 26.17 & 21.62 & 29.19 & 17.74 & 31.55 \\ 
$W_{\odot}^{\mbox{\tiny{RSR}}}$ [${\rm km \, s}^{-1}$] & & 7.25 & 7.26 & 6.79 & 7.70 & 6.35 & 8.14 \\
\hline
$\mu_l$ [${\rm mas \, yr}^{-1}$] & $(-6.379 \pm 0.026)$ \citep{Reid:2004rd} & -6.376 & -6.374 & -6.401 & -6.350 & -6.426 & -6.326 \\
$A+B$ [${\rm km \, s}^{-1} \, {\rm kpc}^{-1}$] & $(0.18 \pm 0.47)$ \citep{Catena:2009mf} & 0.34 & 0.29 & -0.08 & 0.75 & -0.49 & 1.15 \\
$\Sigma(R_0,1.1 \mbox{ kpc}$ [$M_\odot$ ${\rm pc}^{-2}$] & $(71.0 \pm 6.0)$ \citep{Kuijken:1989} & 73.3 & 72.4 & 69.1 & 77.4 & 65.1 & 81.5 \\
$\Sigma_\ast(R_0)$ [$M_\odot \, {\rm pc}^{-2}$] & $(48.0 \pm 8.0)$ \citep{Kuijken:1991} & 44.3 & 43.8 & 39.5 & 49.1 & 34.8 & 53.9 \\
$\rho_0$ [${\rm GeV \, cm}^{-3}$] & & 0.41 & 0.41 & 0.38 & 0.44 & 0.36 & 0.47 \\
$\Theta_0$ [${\rm km \, s}^{-1}$] & & 220.38 & 221.60 & 211.45 & 229.36 & 202.50 & 237.47 \\
$v_{\rm esc}$ [$\ast$] [${\rm km \, s}^{-1}$] & $492-587$ \citep{Piffl:2013mla} & 526 & 483 & 481 & 571 & 455 & 645 \\
$r_{\rm vir}$ [$\ast$] [${\rm kpc}$] & & 288.51 & 246.95 & 247.09 & 330.14 & 229.71 & 406.52 \\
$c_{\rm vir}$ [$\ast$] & $(9.5 \pm 2.6)$ \citep{BoylanKolchin:2009an} & 19.2 & 24.7 & 13.4 & 25.2 & 9.7 & 30.4 \\
$M_\chi$ [$\ast$] [$M_\odot$] & & $10^{12.11}$ & $10^{11.93}$ & $10^{11.93} $ & $10^{12.30}$ & $10^{11.83}$ & $10^{12.58}$ \\
$M_{\rm d}$ [$\ast$] [$M_\odot$] & & $10^{10.67}$ & $10^{10.65}$ & $10^{10.61}$ & $10^{10.73}$ & $10^{10.56}$ & $10^{10.77}$ \\
$M_{\rm bb}$ [$\ast$] [$M_\odot$] & & $10^{7.77}$ & $10^{7.11}$ & $10^{6.54}$ & $10^{9.05}$ & $10^{6.02}$ & $10^{9.78}$ \\
$M_{\chi}/(M_{\rm d}+M_{\rm bb}+M_{\rm gas})$ $[\ast]$ & 28.22 (Eq.~\ref{eqn:ratio}) & 30.03 & 19.01 & 18.97 & 39.78 & 15.69 & 73.13 \\
$M_{\mbox{\tiny{tot}}}(<50 \mbox{ kpc})$ $[\ast]$ [$10^{11} M_\odot$] & $5.4^{+0.2}_{-3.6}$ \citep{Wilkinson:1999hf} & 4.9 & 4.0 & 4.0 & 5.7 & 3.5 & 7.0 \\
$M_{\mbox{\tiny{tot}}}$ $[\ast]$ [$10^{12} M_\odot$] & $1.9^{+3.6}_{-0.5}$ \citep{Wilkinson:1999hf} & 1.5 & 0.9 & 0.9 & 2.1 & 0.7 & 3.8 \\
$M_{\chi}(<R_0)$ [$\ast$] [$M_\odot$] & & $10^{10.69}$ & $10^{10.74}$ & $10^{10.60}$ & $10^{11.12} $ & $10^{10.51}$ & $10^{11.12}$ \\
$M_{\rm d}(<R_0)$ [$\ast$] [$M_\odot$] & & $10^{10.60}$ & $10^{10.57}$ & $10^{10.53}$ & $10^{10.66}$ & $10^{10.46}$ & $10^{10.71}$ \\
\hline
\end{tabular}
\caption{\label{tab:results_NFW} Probability distributions of the parameters in the scan, and other relevant quantities, for the case of a NFW DM halo. The scanned parameters are defined in Tab. \ref{tab:parameters}, while $\mu_l$ is the proper motion of Sgr A$^\star$ and $A+B$ is the sum of the Oort constants (see Sec. \ref{sec:circular_velocity}), $\Sigma(R_0,1.1 \mbox{ kpc})$ is the local (i.e. at $r=R_0$) total surface density within 1.1 kpc of the Galactic plane and $\Sigma_\ast(R_0)$ is the local surface density of the visible component only (see Sec. \ref{sec:local_surface_density}). $\rho_0$ and $\Theta_0$ are the local DM density and circular velocity, respectively. $v_{\rm esc}$ is the local escape speed and $r_{\rm vir}$ ($c$) is the virial radius (concentration). $M_\chi$, $M_{\rm d}$ and $M_{\rm bb}$ are the total (within the virial radius) masses for the DM, disk and bulge/bar components respectively, $M_{\chi}/(M_{\rm d}+M_{\rm bb}+M_{\rm gas})$ is the ratio of DM to baryons, while $M_{\chi}(<50 \mbox{ kpc})$ and $M_{\mbox{\tiny{tot}}}$ are the total mass enclosed within 50 kpc and the virial radius, respectively. Finally, $M_{\chi}(<R_0)$ and $M_{\rm d}(<R_0)$ are the DM and disk mass enclosed within $R_0$ respectively. The second column shows the experimentally measured value (when available). Quantities labelled with $\ast$ are not included as constraints in the scan. The third  and fourth columns contain the posterior mean and best-fit point. The fifth and sixth (seventh and eighth) columns indicate the lower and upper edges of the 68\% (95\%) credible interval.}
\end{center}
\end{table*}

\begin{table*}
\begin{center}
\begin{tabular}{cccccccc}
\multicolumn{8}{c}{\bf Einasto} \\
\hline
{\bf Name} & {\bf Measured value} & {\bf mean} & {\bf best-fit} & {\bf lower 68\%} & {\bf upper 68\%} & {\bf lower 95\%} & {\bf upper 95\%} \\
\hline
$\rho_{\rm s}$ [$M_\odot \, {\rm pc}^{-3}$] & & $10^{-2.28}$ & $10^{-2.24}$ & $10^{-2.61}$ & $10^{-1.94}$ & $10^{-2.91}$ & $10^{-1.68}$ \\
$r_{\rm s}$ [${\rm kpc}$] & & 13.14 & 11.25 & 8.09 & 18.15 & 6.14 & 26.77 \\
$\alpha$ & & 0.21 & 0.20 & 0.13 & 0.30 & 0.11 & 0.38 \\
$R_{\rm d}$ [${\rm kpc}$] & & 2.66 & 2.62 & 2.31 & 3.03 & 2.08 & 3.51 \\
$\Sigma_{\rm d}$ [$M_\odot \, {\rm pc}^{-2}$] & & $10^{2.99}$ & $10^{3.01}$ & $10^{2.83}$ & $10^{3.14}$ & $10^{2.66}$ & $10^{3.25}$ \\
$\rho_{\rm bb}(0)$ [$M_\odot \, {\rm pc}^{-3}$] & & $10^{-2.12}$ & $10^{-3.68}$ & $10^{-3.38}$ & $10^{-0.83}$ & $10^{-3.89}$ & $10^{-0.14}$ \\
$\beta_\ast$ & & -0.35 & -0.30 & -0.64 & -0.07 & -0.89 & 0.17 \\
$R_0$ [${\rm pc}$] & & 8.18 & 8.19 & 7.90 & 8.46 & 7.61 & 8.71 \\
$U_{\odot}^{\mbox{\tiny{RSR}}}$ [${\rm km \, s}^{-1}$] & & 11.09 & 11.02 & 10.17 & 12.01 & 9.27 & 12.91 \\ 
$V_{\odot}^{\mbox{\tiny{RSR}}}$ [${\rm km \, s}^{-1}$] & & 25.33 & 25.77 & 21.44 & 29.13 & 17.51 & 31.53 \\ 
$W_{\odot}^{\mbox{\tiny{RSR}}}$ [${\rm km \,s}^{-1}$] & & 7.25 & 7.28 & 6.79 & 7.70 & 6.34 & 8.15 \\
\hline
$\mu_l$ [${\rm mas \, yr}^{-1}$] & $(-6.379 \pm 0.026)$ \citep{Reid:2004rd} & -6.376 & -6.373 & -6.402 & -6.350 & -6.426 & -6.326 \\
$A+B$ [${\rm km \, s}^{-1} \, {\rm kpc}^{-1}$] & $(0.18 \pm 0.47)$ \citep{Catena:2009mf} & 0.34 & 0.25 & -0.08 & 0.75 & -0.49 & 1.16 \\
$\Sigma(R_0,1.1 \mbox{ kpc}$ [$M_\odot \, {\rm pc}^{-2}$] & $(71.0 \pm 6.0)$ \citep{Kuijken:1989} & 74.2 & 73.8 & 69.8 & 78.5 & 65.7 & 83.1 \\
$\Sigma_\ast(R_0)$ [$M_\odot \, {\rm pc}^{-2}$] & $(48.0 \pm 8.0)$ \cite{Kuijken:1991} & 42.8 & 44.6 & 37.7 & 47.9 & 32.5 & 52.7 \\
$\rho_0$ [${\rm GeV \, cm}^{-3}$] & & 0.42 & 0.41 & 0.39 & 0.45 & 0.36 & 0.48 \\
$\Theta_0$ [${\rm km \, s}^{-1}$] & & 221.81 & 221.62 & 212.98 & 230.66 & 203.98 & 239.04 \\
$v_{\rm esc}$ [$\ast$] [${\rm km \, s}^{-1}$] & $492-587$ \citep{Piffl:2013mla} & 479 & 470 & 445 & 513 & 420 & 549 \\
$r_{\rm vir}$ [$\ast$] [${\rm kpc}$] & & 256.75 & 246.87 & 211.14 & 302.07 & 183.81 & 350.82 \\
$c_{\rm vir}$ [$\ast$] & $(9.5 \pm 2.6)$ \citep{BoylanKolchin:2009an} & 21.7 & 21.9 & 15.5 & 28.1 & 12.38 & 36.15 \\
$M_\chi$ [$\ast$] [$M_\odot$] & & $10^{11.96}$ & $10^{11.93}$ & $10^{11.72} $ & $10^{12.19}$ & $10^{11.54}$ & $10^{12.38}$ \\
$M_{\rm d}$ [$\ast$] [$M_\odot$] & & $10^{10.63}$ & $10^{10.64}$ & $10^{10.56}$ & $10^{10.69}$ & $10^{10.49}$ & $10^{10.75}$ \\
$M_{\rm bb}$ [$\ast$] [$M_\odot$] & & $10^{7.78}$ & $10^{6.23}$ & $10^{6.53}$ & $10^{9.08}$ & $10^{6.02}$ & $10^{9.77}$ \\
$M_{\mbox{\tiny{tot}}}/(M_{\rm d}+M_{\rm bb}+M_{\rm gas})$ $[\ast]$ & 29.01 (Eq.~\ref{eqn:ratio}) & 18.90 & 16.10 & 11.22 & 26.42 & 7.82 & 38.17 \\
$M_{\mbox{\tiny{tot}}}(<50 \mbox{ kpc})$ $[\ast]$ [$10^{11} M_\odot$] & $(5.4 \pm 0.2)$ \citep{Wilkinson:1999hf} & 4.4 & 4.2 & 3.7 & 5.2 & 3.1 & 5.9 \\
$M_{\rm tot}$ $[\ast]$ [$10^{12} M_\odot$] & $1.9^{+3.6}_{-0.5}$ \citep{Wilkinson:1999hf} & 0.9 & 0.7 & 0.5 & 1.2 & 0.4 & 1.7 \\
$M_{\chi}(<R_0)$ [$\ast$] [$M_\odot$] & & $10^{10.74}$ & $10^{10.74}$ & $10^{10.66}$ & $10^{10.82} $ & $10^{10.57}$ & $10^{10.89}$ \\
$M_{\rm d}(<R_0)$ [$\ast$] [$M_\odot$] & & $10^{10.53}$ & $10^{10.56}$ & $10^{10.45}$ & $10^{10.62}$ & $10^{10.36}$ & $10^{10.69}$ \\
\hline
\end{tabular}
\caption{\label{tab:results_Einasto} Same as Tab. \ref{tab:results_NFW} but for the case of an Einasto DM halo.}
\end{center}
\end{table*}

\begin{table*}
\begin{center}
\begin{tabular}{cccccccc}
\multicolumn{8}{c}{\bf Burkert} \\
\hline
{\bf Name} & {\bf Measured value} & {\bf mean} & {\bf best-fit} & {\bf lower 68\%} & {\bf upper 68\%} & {\bf lower 95\%} & {\bf upper 95\%} \\
\hline
$\rho_{\rm s}$ [$M_\odot \, {\rm pc}^{-3}$] & & $10^{-1.04}$ & $10^{-0.89}$ & $10^{-1.38}$ & $10^{-0.72}$ & $10^{-1.61}$ & $10^{-0.53}$ \\
$r_{\rm s}$ [${\rm kpc}$] & & 5.92 & 4.55 & 3.91 & 8.10 & 3.26 & 11.67 \\
$R_{\rm d}$ [${\rm kpc}$] & & 2.58 & 2.76 & 2.23 & 2.95 & 2.06 & 3.47 \\
$\Sigma_{\rm d}$ [$M_\odot \, {\rm pc}^{-2}$] & & $10^{3.11}$ & $10^{3.00}$ & $10^{2.94}$ & $10^{3.28}$ & $10^{2.75}$ & $10^{3.35}$ \\
$\rho_{\rm bb}(0)$ [$M_\odot \, {\rm pc}^{-3}$] & & $10^{-2.17}$ & $10^{-0.37}$ & $10^{-3.40}$ & $10^{-0.88}$ & $10^{-3.89}$ & $10^{-0.20}$ \\
$\beta_\ast$ & & -0.17 & -0.09 & -0.46 & 0.11 & -0.79 & 0.32 \\
$R_0$ [${\rm kpc}$] & & 8.23 & 8.25 & 7.93 & 8.52 & 7.62 & 8.79 \\
$U_{\odot}^{\mbox{\tiny{RSR}}}$ [${\rm km \, s}^{-1}$] & & 11.09 & 11.13 & 10.17 & 12.00 & 9.27 & 12.91 \\ 
$V_{\odot}^{\mbox{\tiny{RSR}}}$ [${\rm km \, s}^{-1}$] & & 24.26 & 24.95 & 20.14 & 28.32 & 16.14 & 31.19 \\ 
$W_{\odot}^{\mbox{\tiny{RSR}}}$ [${\rm km \, s}^{-1}$] & & 7.25 & 7.18 & 6.80 & 7.70 & 6.35 & 8.14 \\
\hline
$\mu_l$ [${\rm mas \, yr}^{-1}$] & $(-6.379 \pm 0.026)$ \citep{Reid:2004rd} & -6.377 & -6.379 & -6.402 & -6.352 & -6.426 & -6.328 \\
$A+B$ [${\rm km \, s}^{-1} \, {\rm kpc}^{-1}$] & $(0.18 \pm 0.47)$ \citep{Catena:2009mf} & 0.41 & 0.33 & 0.30 & 0.83 & -0.40 & 1.25 \\
$\Sigma(R_0,1.1 \mbox{ kpc}$ [$M_\odot \, {\rm pc}^{-2}$] & $(71.0 \pm 6.0)$ \citep{Kuijken:1989} & 68.2 & 67.9 & 64.0 & 72.4 & 59.8 & 76.6 \\
$\Sigma_\ast(R_0)$ [$M_\odot \, {\rm pc}^{-2}$] & $(48.0 \pm 8.0)$ \citep{Kuijken:1991} & 50.7 & 50.4 & 46.1 & 55.4 & 41.7 & 60.0 \\
$\rho_0$ [${\rm GeV\, cm}^{-3}$] & & 0.41 & 0.41 & 0.38 & 0.44 & 0.36 & 0.47 \\
$\Theta_0$ [${\rm km \, s}^{-1}$] & & 224.50 & 224.48 & 214.71 & 234.26 & 204.56 & 242.69 \\
$v_{\rm esc}$ [$\ast$] [${\rm km \, s}^{-1}$] & $492-587$ \citep{Piffl:2013mla} & 461 & 440 & 433 & 489 & 413 & 538 \\
$r_{\rm vir}$ [$\ast$] [${\rm kpc}$] & & 217.93 & 201.88 & 198.06 & 238.33 & 188.17 & 281.27 \\
$c_{\rm vir}$ [$\ast$] & $(9.5 \pm 2.6)$ \citep{BoylanKolchin:2009an} & 40.2 & 44.4 & 29.2 & 51.5 & 23.8 & 60.6 \\
$M_\chi$ [$\ast$] [$M_\odot$] & & $10^{11.97}$ & $10^{11.88}$ & $10^{11.85} $ & $10^{12.10}$ & $10^{11.78}$ & $10^{12.31}$ \\
$M_{\rm d}$ [$\ast$] [$M_\odot$] & & $10^{10.72}$ & $10^{10.68}$ & $10^{10.66}$ & $10^{10.79}$ & $10^{10.59}$ & $10^{10.84}$ \\
$M_{\rm bb}$ [$\ast$][$M_\odot$] & & $10^{7.73}$ & $10^{9.52}$ & $10^{6.50}$ & $10^{9.02}$ & $10^{6.01}$ & $10^{9.71}$ \\
$M_{\chi}/(M_{\rm d}+M_{\rm bb}+M_{\rm gas})$ $[\ast]$ & 28.04 (Eq.~\ref{eqn:ratio}) & 10.87 & 8.94 & 8.70 & 12.56 & 7.61 & 19.85 \\
$M_{\mbox{\tiny{tot}}}(<50 \mbox{ kpc})$ $[\ast]$ [$10^{11} M_\odot$] & $(5.4 \pm 0.2)$ \citep{Wilkinson:1999hf} & 3.6 & 3.1 & 3.0 & 4.2 & 2.7 & 5.3 \\
$M_{\mbox{\tiny{tot}}}$ $[\ast]$ [$10^{12} M_\odot$] & $1.9^{+3.6}_{-0.5}$ \citep{Wilkinson:1999hf} & 0.7 & 0.5 & 0.5 & 0.8 & 0.4 & 1.3 \\
$M_{\chi}(<R_0)$ [$\ast$] [$M_\odot$] & & $10^{10.68}$ & $10^{10.72}$ & $10^{10.54}$ & $10^{10.81} $ & $10^{10.41}$ & $10^{10.89}$ \\
$M_{\rm d}(<R_0)$ [$\ast$] [$M_\odot$] & & $10^{10.64}$ & $10^{10.58}$ & $10^{10.55}$ & $10^{10.73}$ & $10^{10.45}$ & $10^{10.79}$ \\
\hline
\end{tabular}
\caption{\label{tab:results_Burkert} Same as Tab. \ref{tab:results_NFW} but for the case of a Burkert DM halo.}
\end{center}
\end{table*}

The three mass models associated with the different DM halos share the 
following properties: in the $(\rho_s,r_s)$ plane, lines of constant 
$M_\chi(<R_0)$, the DM mass within the solar radius, are approximately parallel 
to lines with constant $\Sigma_\chi(R_0,1.1 \mbox{ kpc})$, as well as lines 
with constant $\rho_0$. Thus, imposing the constraints on 
$\Sigma(R_0,1.1 \mbox{ kpc})$ and $\Sigma_\ast(R_0)$ (fixing, indirectly, the 
local DM surface density), will select one of those lines (determining one 
degenerate direction in the plane $(\rho_s,r_s)$) and will translate into a 
determination of $M_\chi(<R_0)$ and $\rho_0$. The degeneracy is broken when we 
include the information on the velocity dispersion, which fixes the DM mass 
inside larger radii.
On the other hand, in the $(\Sigma_d,R_d)$ plane, the lines of constant 
$\Sigma_{\rm d}(R_0,1.1 \mbox{ kpc})$ are not parallel to those with constant
$M_{\rm d}(<R_0)$, so that the information on $\Sigma_\ast(R_0)$ and on $\Theta_0$
(constraining the total amount of matter within $R_0$) act in a complementary
way. The allowed region in the  $(\Sigma_d,R_d)$ plane gets slightly 
larger when the terminal velocity data are included, since they prefer a 
less dense disk and a balance has to be made between the constraints on 
$\Sigma_\ast(R_0)$ and $\Theta_0$.

Microlensing and the proper motions of masers do not introduce significant 
addition information to the determination of the mass model. In the case of 
the microlensing data the baryonic component is already well determined by 
the information on the local surface density and the rotation curve. While, 
for the masers, since we are marginalizing over the velocity of the peculiar 
motion of each maser, the scan practically has the effect of determining 
which values of $(U_s,V_s,W_s)$ (for each object) provide a good fit to the 
data, given a rotation velocity fixed by the constraints on the local surface 
density and the rotation curve. This result was already discussed by 
Ref. \citep{Bovy:2009dr}.

The gravitational potential is dominated by the disk for $R<R_0$, while the DM 
halo only becomes important around the Sun's radius. The bulge/bar is always
subdominant: the probability distribution for $\rho_{\rm bb}(0)$ is almost flat
for values smaller than approximately 1 $M_\odot \, \mbox{pc}^{-3}$ and then goes
rapidly to zero. This means that one should regard the upper end of the
68\% and 95\% credible contours for $\rho_{\rm bb}$ as upper limits, while the
lower end is practically determined by the range of values scanned.
 
\begin{figure*}
\includegraphics[width=0.32\textwidth]{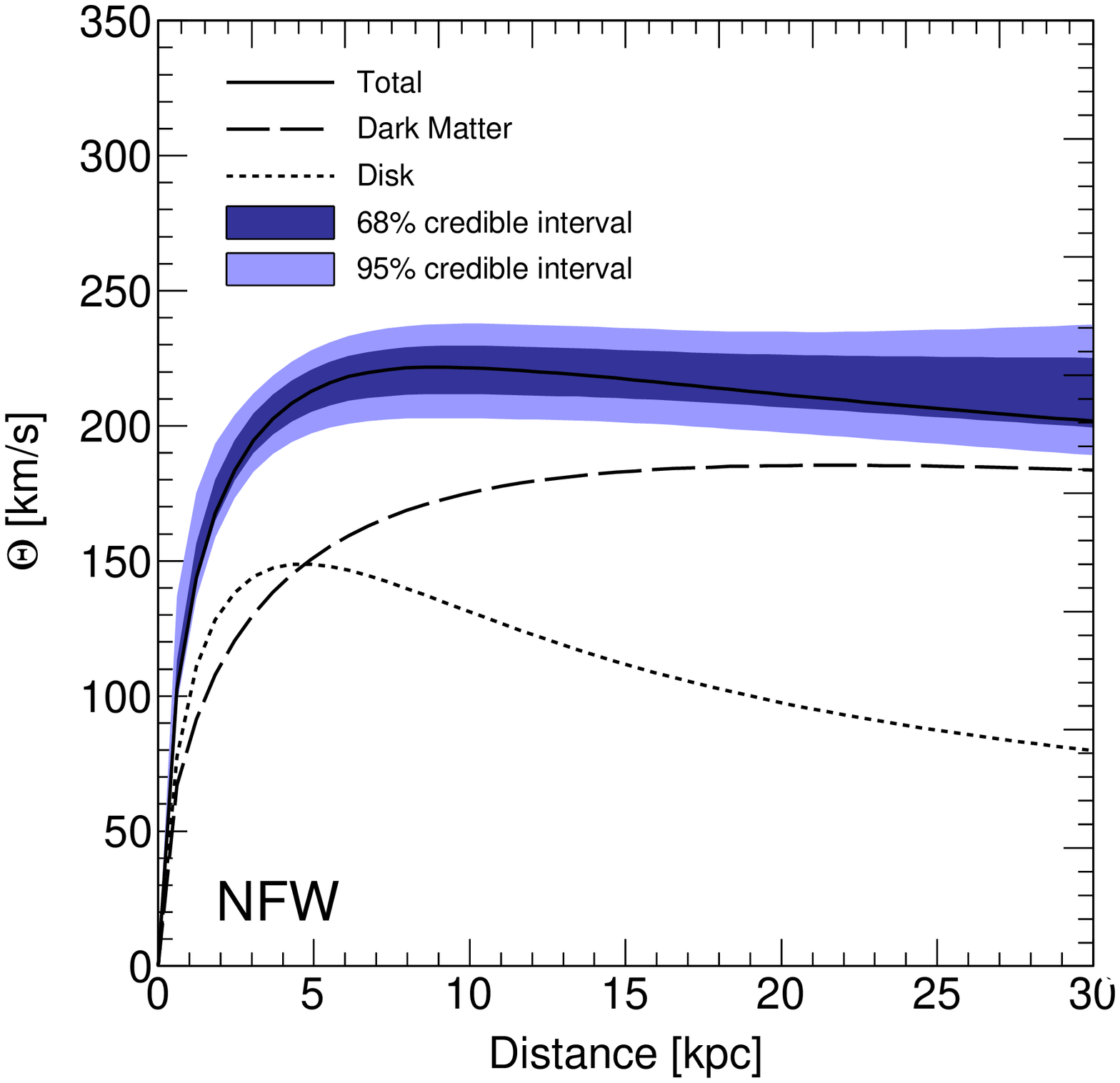}
\includegraphics[width=0.32\textwidth]{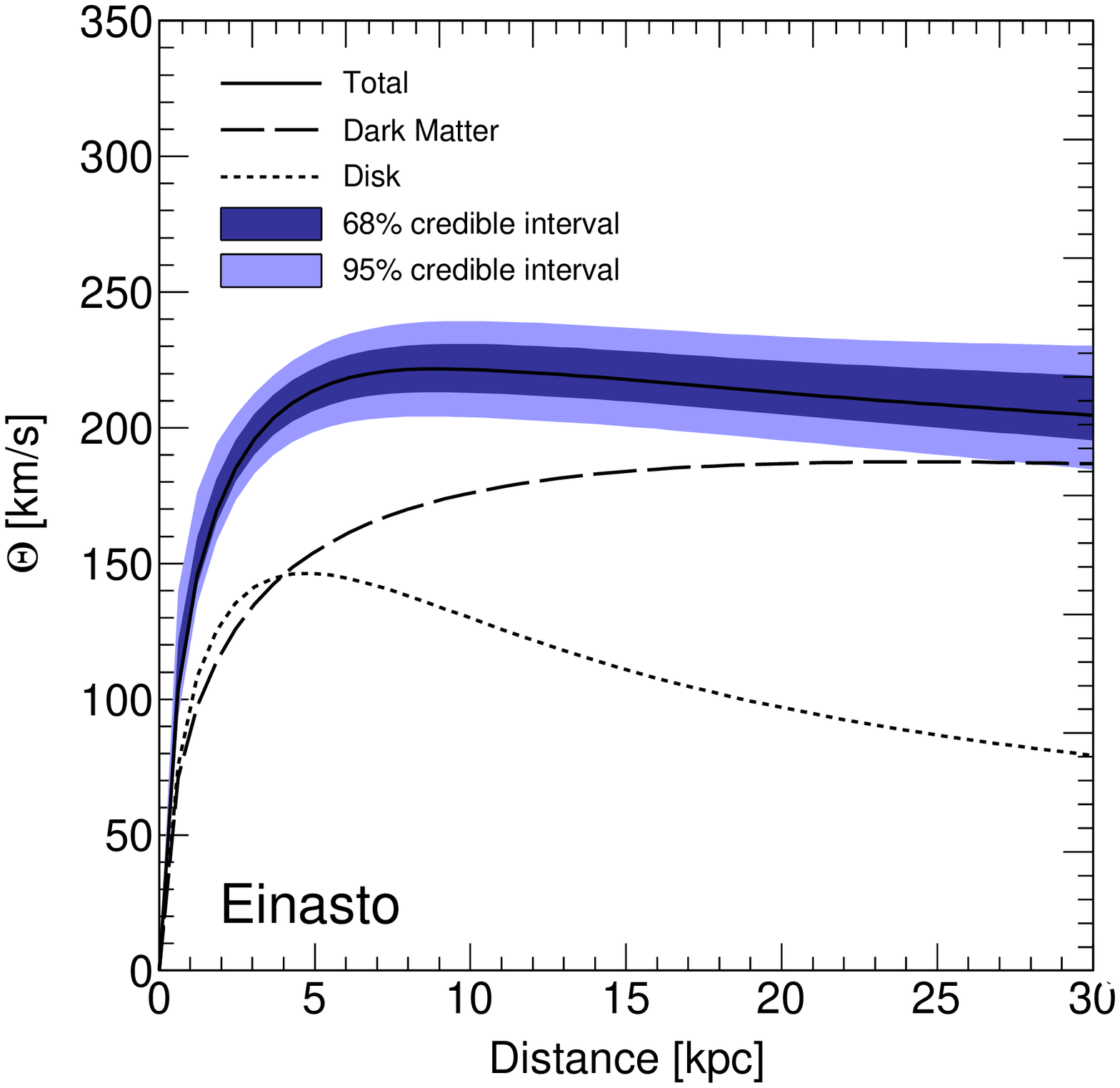}
\includegraphics[width=0.32\textwidth]{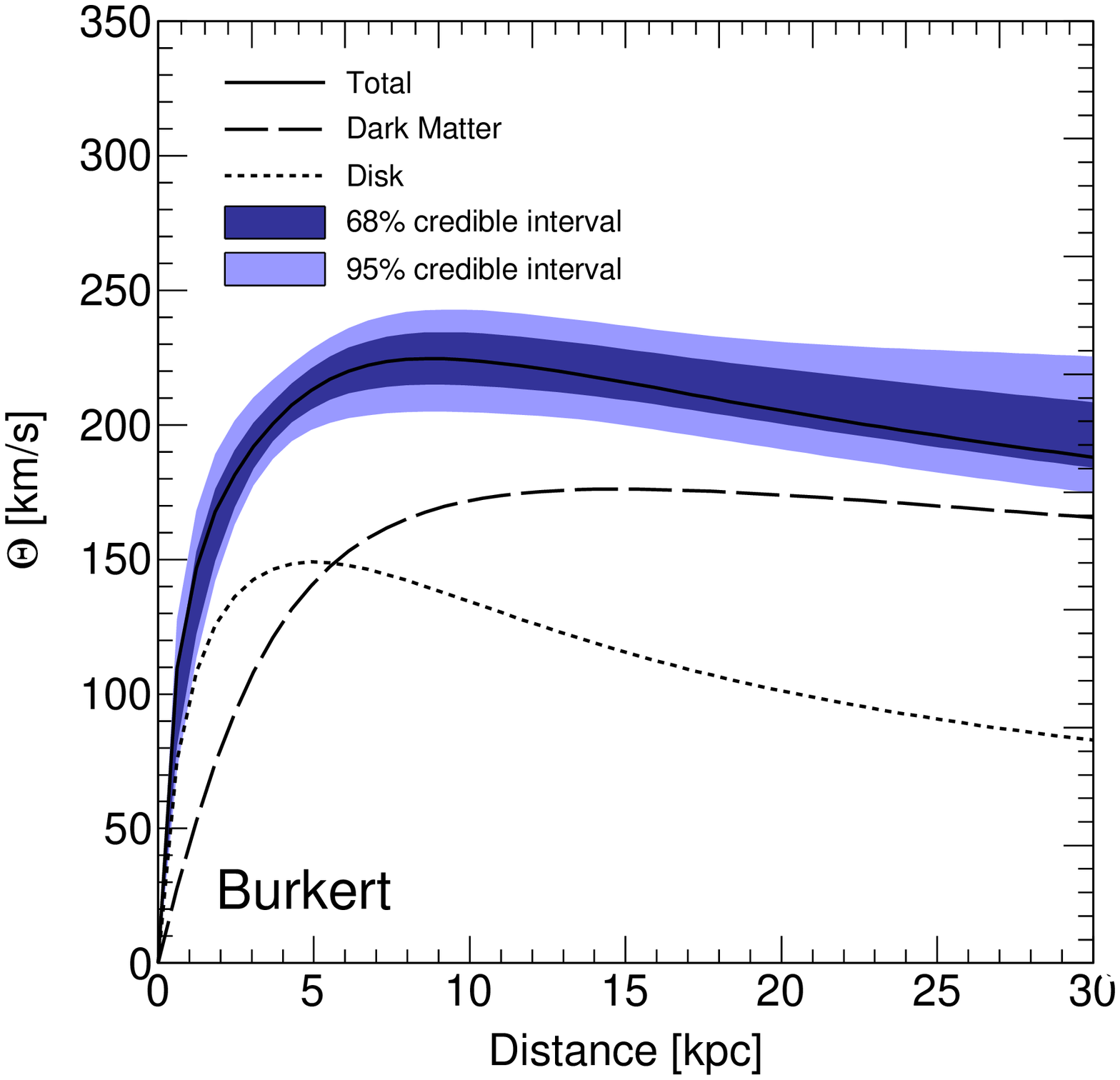}
\caption{\label{fig:rotation_curves} The dark (light) blue band shows the 68\% (95\%) credible region for the reconstruction of the rotation curve $\Theta(r)$ of the MW for, from left to right, the NFW, Einasto and Burkert profiles for the DM halo. The solid black line shows the best-fit rotation curve, while the dashed (dotted) line indicates the contribution of the DM halo (disk).}
\end{figure*}

Fig. \ref{fig:rotation_curves} shows the posterior pdf for the rotation curve
$\Theta(r)$. The dark and light blue band indicate the 68\% and 95\% credible 
regions, respectively. The solid black line corresponds to the best-fit point
and the fact that it falls within the 68\% credible interval reflects
the similar shape of the pdf and PL contours. The dashed (dotted) line
shows the contribution of the DM halo (disk) to the MW rotation curve.

No tension is present between the results of the scans and the experimental
data considered. At the best-fit point (independently of which profile is 
used for the DM halo) the $\chi^2$ is dominated by the proper motion of masers,
which is responsible for approximately half of the value of $\chi^2$.

The results of Tab. \ref{tab:results_NFW} for the case of a NFW DM halo are
quite similar to the case of an Einasto halo, since the two parametrizations
differ mainly in the inner region, where DM does not dominate the 
gravitational potential. On the other hand, a larger concentration and a
smaller virial mass for the DM halo are obtained for the case of a Burkert 
profile. This is a consequence of the presence of the core: fixing the local 
DM density to a common value for the three halo models (as a result of the 
measurement of the local circular velocity and of the local surface density) 
implies a lower concentration for the Burkert DM halo. We stress that the 
goal of this paper is to infer local quantities (specifically the velocity 
distribution $f(\mathbf{v})$), not to determine whether a cuspy or cored 
profile provides a better fit to the data. Thus we leave a comparison of the 
goodness of fit of the different models for future work.

Our results are similar to those of Ref. \citep{Catena:2009mf}. The main 
difference is our value for the local circular velocity, which is $\sim$ 20 
km/s smaller than theirs, independently of the profile chosen for the DM halo. 
This comes from the fact that Ref. \citep{Catena:2009mf} included a component 
of uncollapsed baryons, which they assumed follow the same density profile as 
the DM halo. It is thus expected that, for approximately the same values of 
other parameters, they find a larger circular velocity\footnote{The large 
value of $\Theta_0$ found in Ref. \citep{Catena:2009mf} is still compatible 
with the measurement of the proper motion of Sgr A$^\ast$ that they (and we) 
use, because Ref. \citep{Catena:2009mf} assumed a lower $V_{\phi,\odot}$.}. This 
is also related to the fact that the bulge/bar component in 
Ref. \citep{Catena:2009mf} plays a significant role in the (very center of 
their) gravitational potential: in our case, the disk and DM components (fixed 
by the constraints on the local surface density) already saturates our 
smaller value of $\Theta_0$ without leaving any room for a significant 
bulge/bar contribution.

Ref. \citep{McMillan:2011wd} also found a large value of $\Theta_0$, similar 
to that quoted in Ref. \citep{Catena:2009mf}. In the case of 
Ref. \citep{McMillan:2011wd} we believe that this is due to their best-fit 
model having a larger contribution from the disk (in particular the thick one) 
which leads to a value of $\Sigma_\ast(R_0)$ larger than the measurement from 
Ref. \citep{Kuijken:1989}, which we use. Note the mass of the bulge in
Ref. \citep{McMillan:2011wd} is fixed since this is one of their observational 
constraints.

Our local DM density is smaller than (but still compatible with) the value 
quoted in Ref. \citep{Nesti:2013uwa}: $\rho_0 \approx 0.5$ GeV  cm$^{-3}$. 
This is a direct consequence, again, of the value assumed for $V_{\phi,\odot}$ 
and the resulting $\Theta_0$.

The local DM density predicted for our mass models is also compatible with
the values obtained by model-independent strategies, e.g., the solution of
the equation of centrifugal equilibrium \citep{Salucci:2010qr} or the Minimal 
Assumption method developed in Ref. \citep{Garbari:2011dh} and applied to the 
study of approximately 2000 K dwarf stars in Ref. \citep{Garbari:2012ff}.

We now consider the quantities marked with $\ast$ in Tabs. 
\ref{tab:results_NFW},\ref{tab:results_Einasto} and \ref{tab:results_Burkert}. 
We do not (for various reasons, as discussed in Sec.~\ref{sec:other})
include these quantities as data in our scans, instead we calculate the 
predicted values for our mass models. Our mass models predict that the mass 
within $50 \, {\rm kpc}$, $M_{\rm tot}(r<50 \mbox{ kpc})$, and the total mass, 
$M_{\rm tot}$, are lower than found in Ref. \citep{Smith:2006ym} and 
Ref. \citep{Wilkinson:1999hf} from the kinematics of satellite galaxies and 
halo stars. We believe this is a consequence of our smaller value of 
$\Theta_0$ (or, equivalently, having assumed a large value of 
$V_{\odot}^{\mbox{\tiny{RSR}}}$).

Note also that our mass model ignores any interaction between the DM halo and 
the baryonic components. Hydrodynamic simulations of MW like galaxies, 
including DM and baryons, e.g. the Eris project \citep{Kuhlen:2013tra}, find 
that baryonic contraction, along with the formation of a DM disc, results in a 
larger local DM density  than in DM-only simulations. However, the DM disc in 
the Eris simulations makes a relatively small contribution to the local DM 
density.

\section{Self-consistent solution for the phase-space density of Dark Matter in the Milky Way halo}
\label{sec:velocity_distribution}
In this section we summarize how we compute the DM phase-space density 
$F(E,L)$ and, consequently, the DM velocity distribution $f(\mathbf{v})$.
We work under the following assumptions:
\begin{itemize}
\item The system is in a steady state. In reality this assumption is unlikely 
to be satisfied completely. Simulated halos contain substructures and the 
high speed tail of the speed distribution has features \citep{Kuhlen:2009vh} 
corresponding to incompletely phase-mixed DM, dubbed `debris flow' 
\citep{Kuhlen:2012fz}. However, at the Solar radius the dominant component 
of the DM distribution is expected to be smooth \citep{Vogelsberger:2008qb}.

\item The system is spherical and anisotropic, so that the phase-space 
distribution function is a function of only the energy and the angular 
momentum, which are integrals of motion. This also implies that the DM 
velocity tensor is diagonal in a spherical coordinate frame (i.e. the mixed 
terms $\sigma_{i,j}$ with $i \neq j$ are zero) and that the DM velocity 
moments satisfy the Jeans equation.

\item The phase-space density is separable in the its two variables: 
$F(E,L)=F_E(E) F_L(L)$. Ref. \citep{Wojtak:2008mg} tested this simplifying 
assumption qualitatively for simulated cluster-sized halos.

\item $F_L(L)$ takes the form
\begin{equation}
F_L(L) = \left( 1 + \frac{L^2}{2L_0^2} \right)^{-\beta_\infty+\beta_0} L^{-2\beta_0} \,,
\label{eqn:F_L}
\end{equation}
where the parameters $\beta_0$, $\beta_\infty$ and $L_0$ are defined in 
Sec. \ref{sec:sampling}.

This ansatz was originally proposed in Ref. \citep{Wojtak:2008mg}, which 
showed that the self-consistent solutions obtained from this assumption 
match the radial dependence of the anisotropy parameter, $\beta(r)$, found in 
simulated cluster-sized halos. The clusters studied in 
Ref. \citep{Wojtak:2008mg} had an isotropic velocity distribution (i.e. 
$\beta \approx 0$) close to their center, $\beta \sim 0.2$ at the scale 
radius, increasing to a value as large as 0.4 at $r \sim 10 \, r_s$. 
Refs. \citep{Ludlow:2010sy,Ludlow:2011cs} found similar behaviour in 
simulations of  MW-like halos. Unlike the cluster-size halos studied by 
Ref. \citep{Wojtak:2008mg}, in galaxy-sized halos $\beta$ may decrease 
beyond $r \sim 5 \, r_s$ \citep{Ludlow:2010sy,Ludlow:2011cs}. However, this 
can be accommodated by the parametrization in Eq.~(\ref{eqn:F_L}). Thus we 
use Eq.~(\ref{eqn:F_L}) to parametrize the $L$-dependent part of the 
phase-space density for the MW DM halo.
\end{itemize}

Once the gravitational potential is fixed, and for a specific choice of the 
three parameters entering in Eq. \ref{eqn:F_L}, it is possible to determine
$F_E(E)$ by inverting the following equation:
\begin{eqnarray}
\rho_\chi(r) & = & \int {\rm d}^3 \, v \, F_E(E) F_L(L) \,,  \nonumber \\
&= & \int {\rm d}^3 \, v \, F_E(E) \left( 1 + \frac{L^2}{2 L_0^2} \right)^{-\beta_\infty + \beta_0} 
L^{-2\beta_0}.
\label{eqn:Volterra}
\end{eqnarray}
This is what guarantees the self-consistency of our solutions, since, for a
given $F_L(L)$, the phase-space density is completely determined by the 
gravitational potential of the system. Thus, reconstructing $F(E,L)$ by 
inverting Eq.~(\ref{eqn:Volterra}) can be considered as an extension of the
Eddington formulism to the case of an anisotropic DM halo\footnote{Note that 
the original Eddington formalism, for a spherical isotropic system, does not 
require any additional parameters and the speed distribution is completely 
determined by the gravitational potential 
\citep{Eddington:1915,Catena:2011kv,Bhattacharjee:2012xm,Pato:2012fw}.}.

The solution has to be determined numerically since Eq.~(\ref{eqn:Volterra})
is a Volterra integral equation. Details can be found in the appendix of 
Ref. \citep{Wojtak:2008mg}. $F_E(E)$ will depend on the gravitational potential
of the system (this becomes evident as soon as one changes the integration 
variables in Eq.~(\ref{eqn:Volterra}) to $E$ and $L$) and on the choice made
for $\beta_0$, $\beta_\infty$ and $L_0$. The astronomical observations listed in
Sec. \ref{sec:constraints} constrain the gravitational potential without, 
however, providing any information on the parameters which appear in 
Eq.~(\ref{eqn:F_L}). Therefore we must marginalize over $\beta_0$, 
$\beta_\infty$ and $L_0$, which can lead to large uncertainties in  $F(E,L)$. 
However, the relevant quantity for direct detection experiments is the local 
speed distribution $f_1(v)$, defined as
\begin{equation}
f_{1}(v) = \int v^2 f(\mathbf{v}) \, {\rm d} \, \Omega_{\mathbf{v}} = \frac{\int v^2 \, F(\mathbf{x}_\odot,\mathbf{v}) \, {\rm d} \,\Omega_{\mathbf{v}}}{\rho_\chi({\mathbf{x}_\odot})},
\end{equation}
and $f_{1}(v)$ turns out to be much less dependent on the form of $F_L(L)$ than 
$F_E(E)$, and therefore it is possible to reconstruct $f_{1}(v)$ with 
reasonable uncertainties even when marginalizing over $\beta_0$, $\beta_{\infty}$ 
and $L_0$.

\begin{figure}
\includegraphics[width=0.45\textwidth]{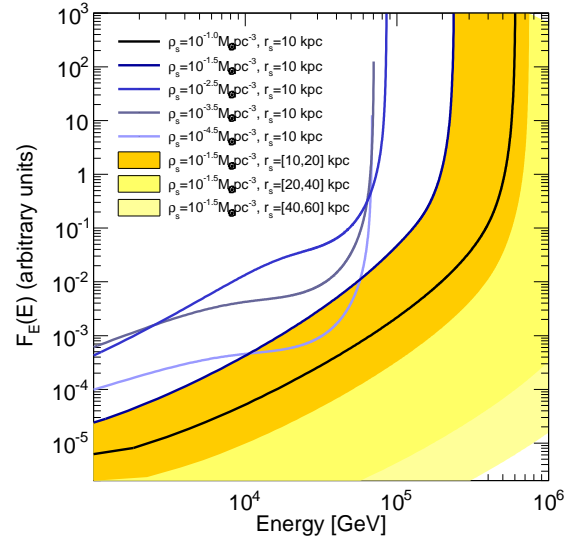}
\caption{\label{fig:F_E} The energy-dependent part of the phase-space density (in arbitrary units). Different shades of blue indicate the dependence on $\rho_{\rm s}$ for fixed $r_{\rm s}$ (the baryonic components of the mass model and the $L$-dependent part of the phase-space density are also fixed). The yellow bands show how $F_E(E)$ changes when $\rho_{s}$ is kept constant and $r_{\rm s}$ is varied.}
\end{figure}

Fig. \ref{fig:F_E} shows how $F_E(E)$ depends on the gravitational potential 
of the system once the parameters entering in $F_L(L)$ are fixed. We only 
consider the effect of changing the characteristic density, $\rho_{\rm s}$, 
and the scale radius, $r_{\rm s}$, of the DM halo. For a given value of 
$r_{\rm s}$, decreasing $\rho_{\rm s}$ corresponds to decreasing the amount of 
DM present in the halo and therefore the maximum value of the gravitational 
potential (i.e. the value at the center of the MW). This is why on 
decreasing the characteristic density, the range of $E$ values over which 
$F_E(E)$ is defined becomes smaller. For $\rho_{\rm s}$ below approximately
$10^{-1.5} M_\odot \mbox{pc}^{-3}$ the baryonic component becomes important in 
particular for the largest values of energy, which correspond to the central 
region of the halo. Once the inner potential is dominated by baryons, the 
behaviour of $F_E(E)$ at high energies is independent of $\rho_{\rm s}$ and the 
only remaining effect is at low energies, where $F_{E}(E)$ decreases as 
$\rho_{\rm s}$ is decreased, simply because there is less DM.

The yellow bands show the behaviour of $F_E(E)$ as the scale radius is varied,
with $\rho_{\rm s}$ kept constant. Increasing $r_{\rm s}$ leads to an
increase in the amount of DM in the MW and therefore $F_{E}(E)$ moves to 
larger values.

\begin{figure*}
\begin{center}
\includegraphics[width=0.49\textwidth]{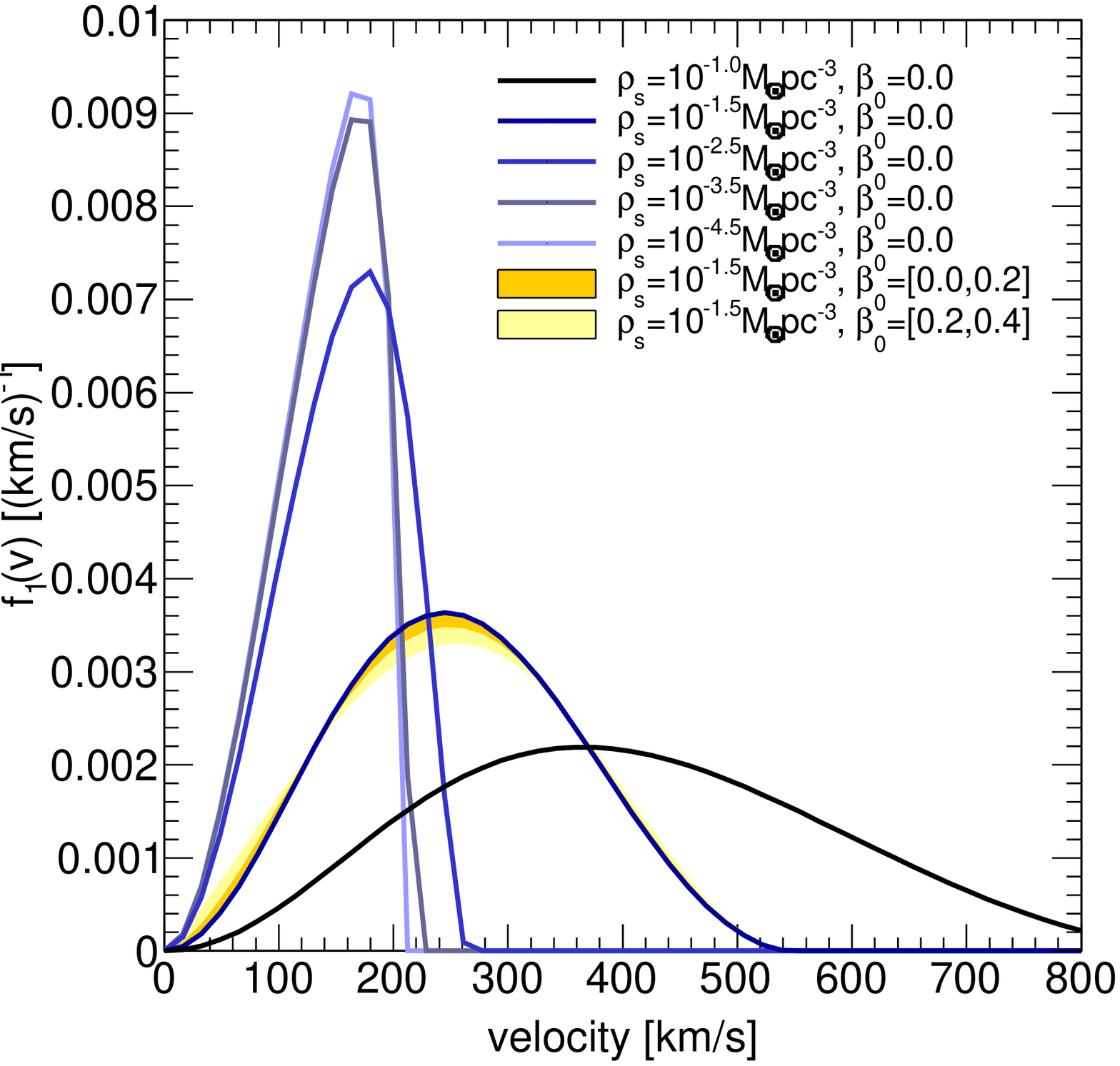}
\includegraphics[width=0.49\textwidth]{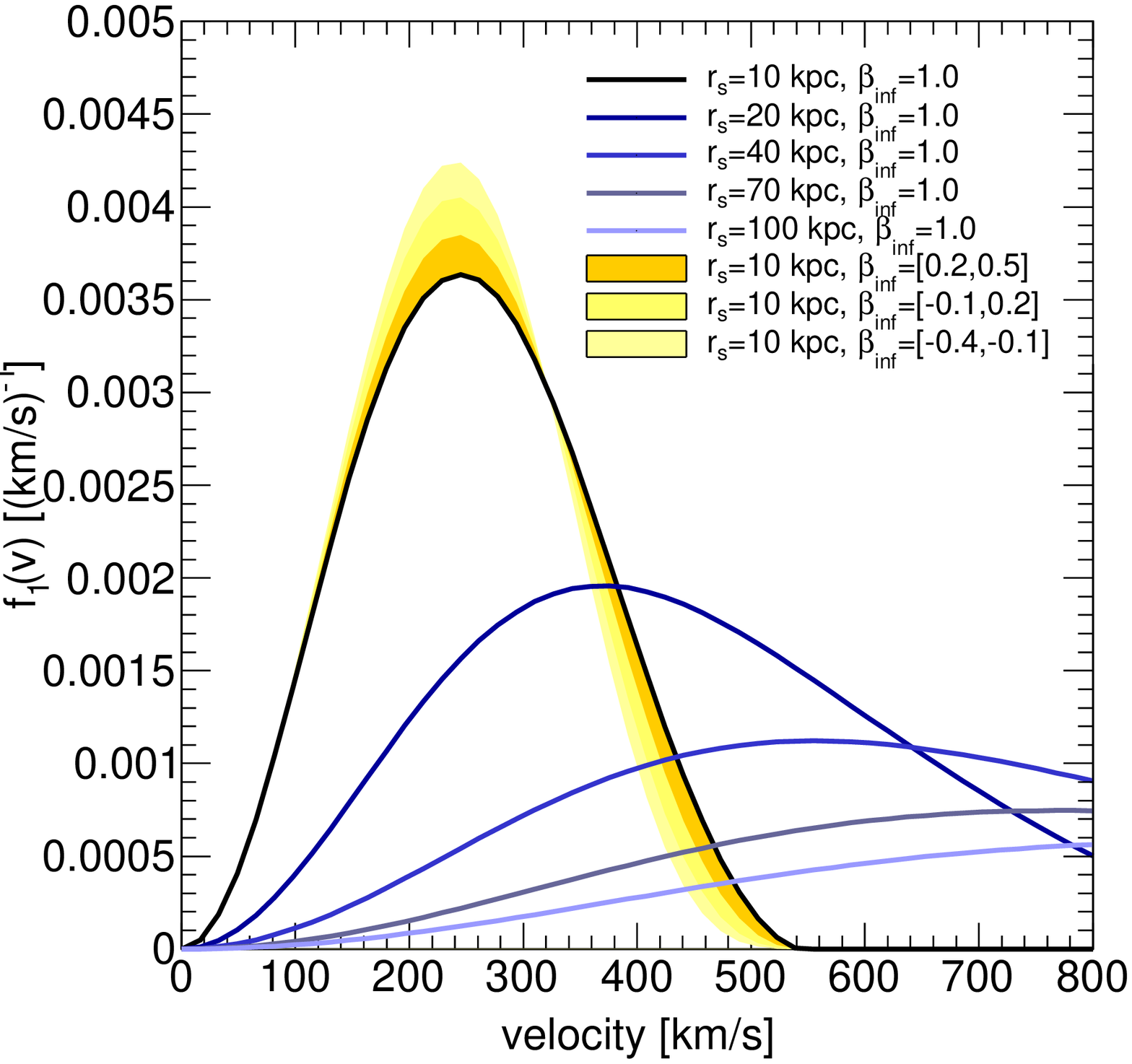}
\includegraphics[width=0.49\textwidth]{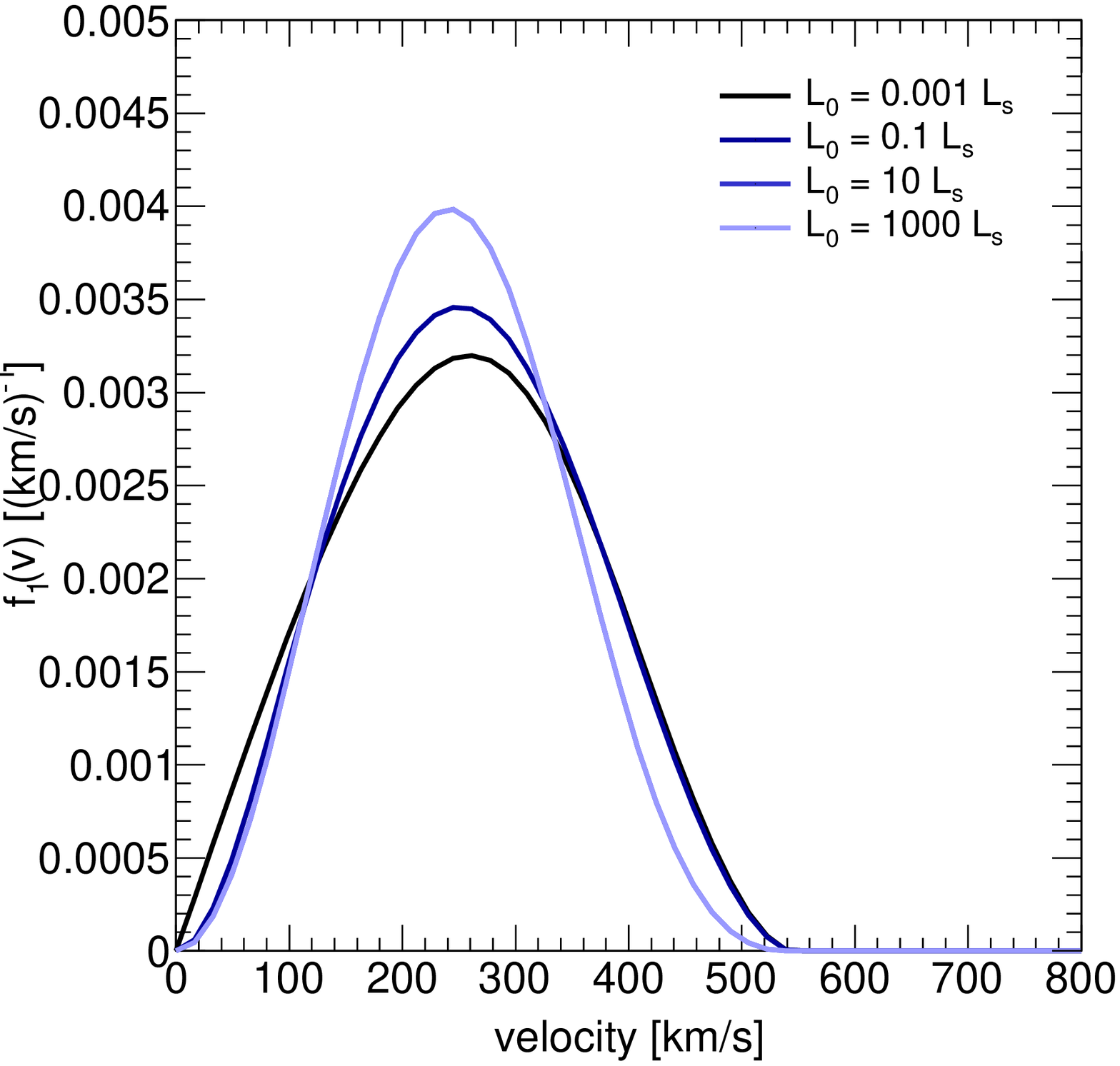}
\caption{\label{fig:fv} Upper left panel:  the $f_1(v)$ distribution for various values of $\rho_{\rm s}$ (all  other parameters are left unchanged). The yellow bands show the effect of changing $\beta_0$. Upper right panel: the same as the left panel but for the dependence on the scale radius (lines) and on $\beta_{\infty}$ (bands). Lower panel: the same as the upper panels but for the dependence on $L_0$. Note that the lines corresponding to $L_0=10 \, L_{\rm s}$ and $L_{0}=1000 \, L_{\rm s}$ coincide.}
\end{center}
\end{figure*}

Similarly Fig. \ref{fig:fv}  shows how $f_{1}(v)$ depends on the gravitational 
potential.
For small values of $\rho_{\rm s}$ the gravitational potential is completely 
determined by the baryonic components and the escape velocity is quite small, 
thus $f_{1}(v)$ goes to zero quite rapidly between 200 and 300 
${\rm  km \, s}^{-1}$. As the DM characteristic density is increased the range 
within which $f_{1}(v)$ is non zero gets larger and, therefore, the peak of the 
distribution is reduced since $f_1(v)$ is normalized to one. 
As in  Fig. \ref{fig:F_E}, increasing the scale radius (with $\rho_{\rm s}$ 
fixed) increases the gravitational potential and the escape velocity, and 
therefore the peak in the speed distribution moves to higher speeds.

The yellow bands show the effect of changing $\beta_0$ and $\beta_{\infty}$.
The effect is small and is localized at small/intermediate 
(intermediate/large) velocities for $\beta_0$ ($\beta_\infty$). Note that the 
gravitational potential is fixed and therefore picking one specific velocity 
completely determines the energy of the DM particle. Small velocities 
correspond to large energies (remember that $E=\Phi(\mathbf{x_\odot})-v^2/2$) 
and to orbits localized close to the center of the halo. On the other hand, 
particles with large velocities will have small energies and will move on 
orbits that can reach large distances. Therefore the low (high)-velocity 
regime changes more as $\beta_0$ ($\beta_{\infty}$) is varied.

For small negative values of $L_{0}$ increasing the transition scale 
has the same effect as decreasing $\beta_0$, since it corresponds to
reducing the portion of the halo with large $\beta$. For the same reason
when the transition happens more or less at $R_0$, the effect of increasing
$L_0$ is the same as increasing $\beta_\infty$. Finally, for large $L_0$, the
velocity distribution becomes independent of $L_0$ since increasing the 
transition scale only affects large radii.

The ranges we take for the parameters defining $F_{L}(L)$ in Fig. \ref{fig:fv} 
approximately match the ranges over which we carry out our scans (see 
Tab. \ref{tab:parameters}). The anisotropy at the center is constrained to be 
smaller than half the central slope of the DM halo profile \citep{An:2005tm}. 
This imposes an upper limit of 0.5 for the case of a NFW halo and only allows 
zero or negative $\beta_0$ for Einasto and Burkert profiles. However the 
halos formed in $N$-body simulations do not have negative $\beta_0$, and 
therefore we only consider $\beta_0 \geq -0.5$. We allow $\beta_\infty$ to vary 
between -0.5 and 1.0, allowing for the possibility of $\beta_\infty < \beta_0$.

\section{Results and discussion}
\label{sec:discussion}
\begin{figure*}
\includegraphics[width=0.32\textwidth]{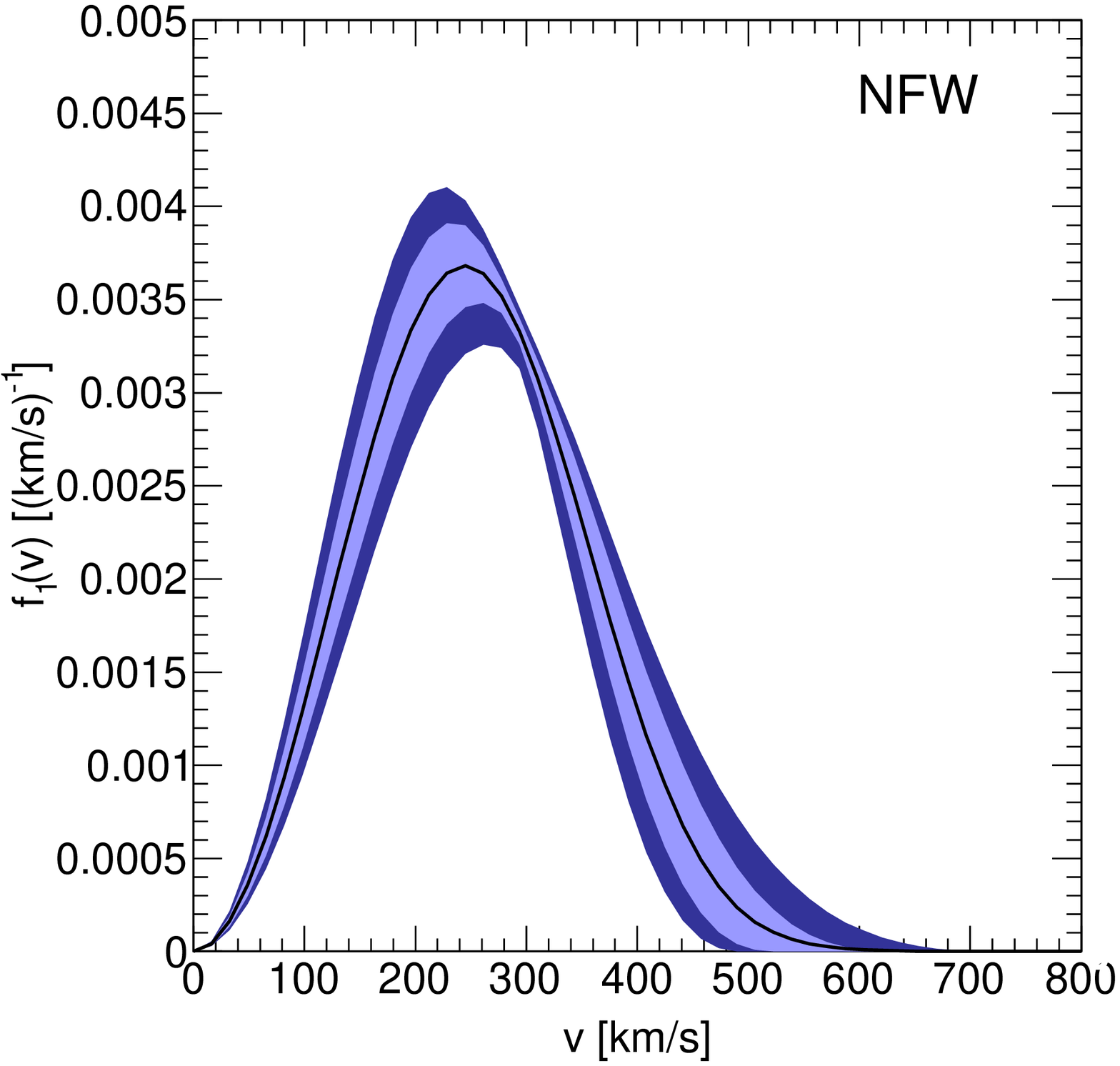}
\includegraphics[width=0.32\textwidth]{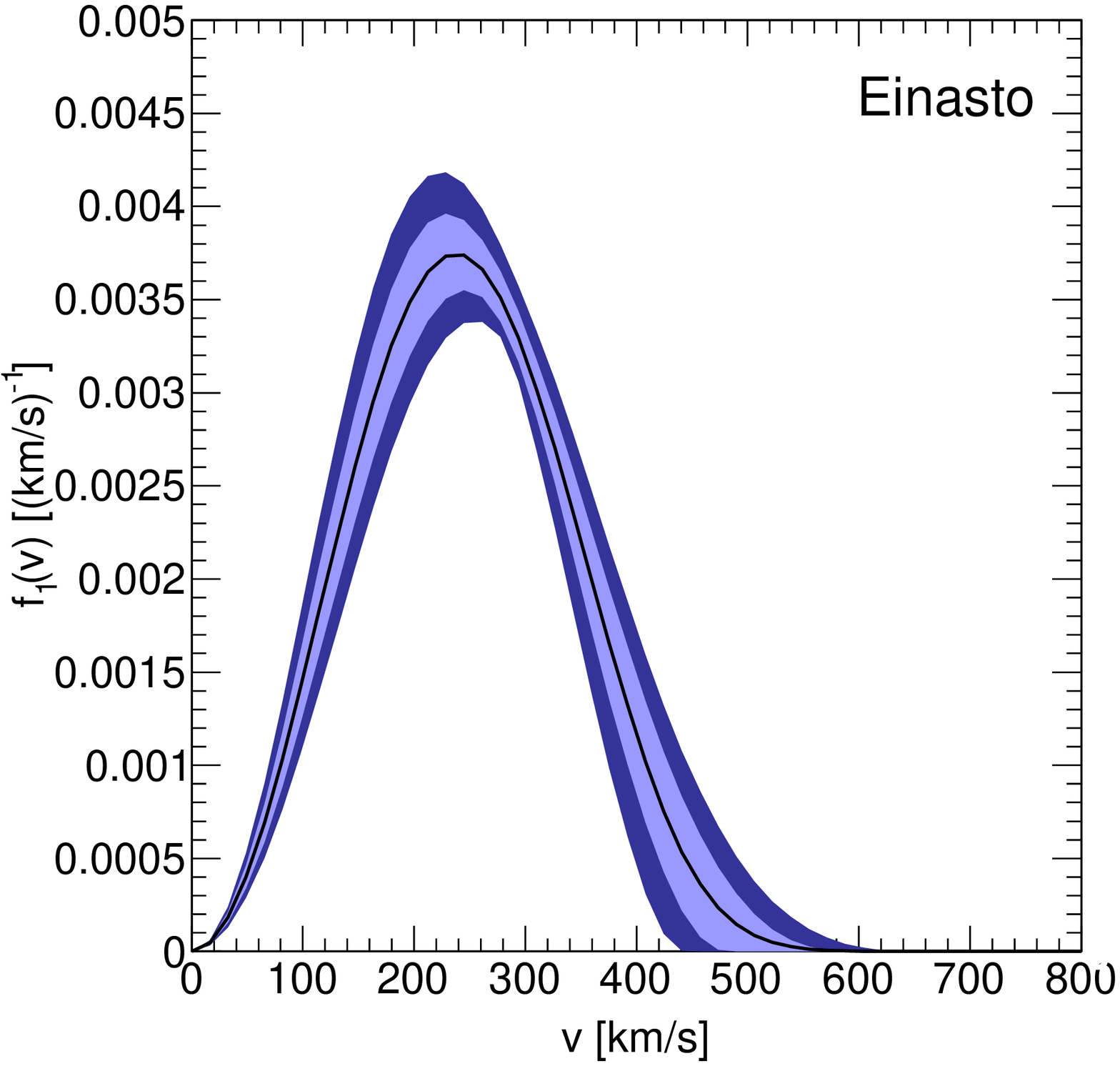}
\includegraphics[width=0.32\textwidth]{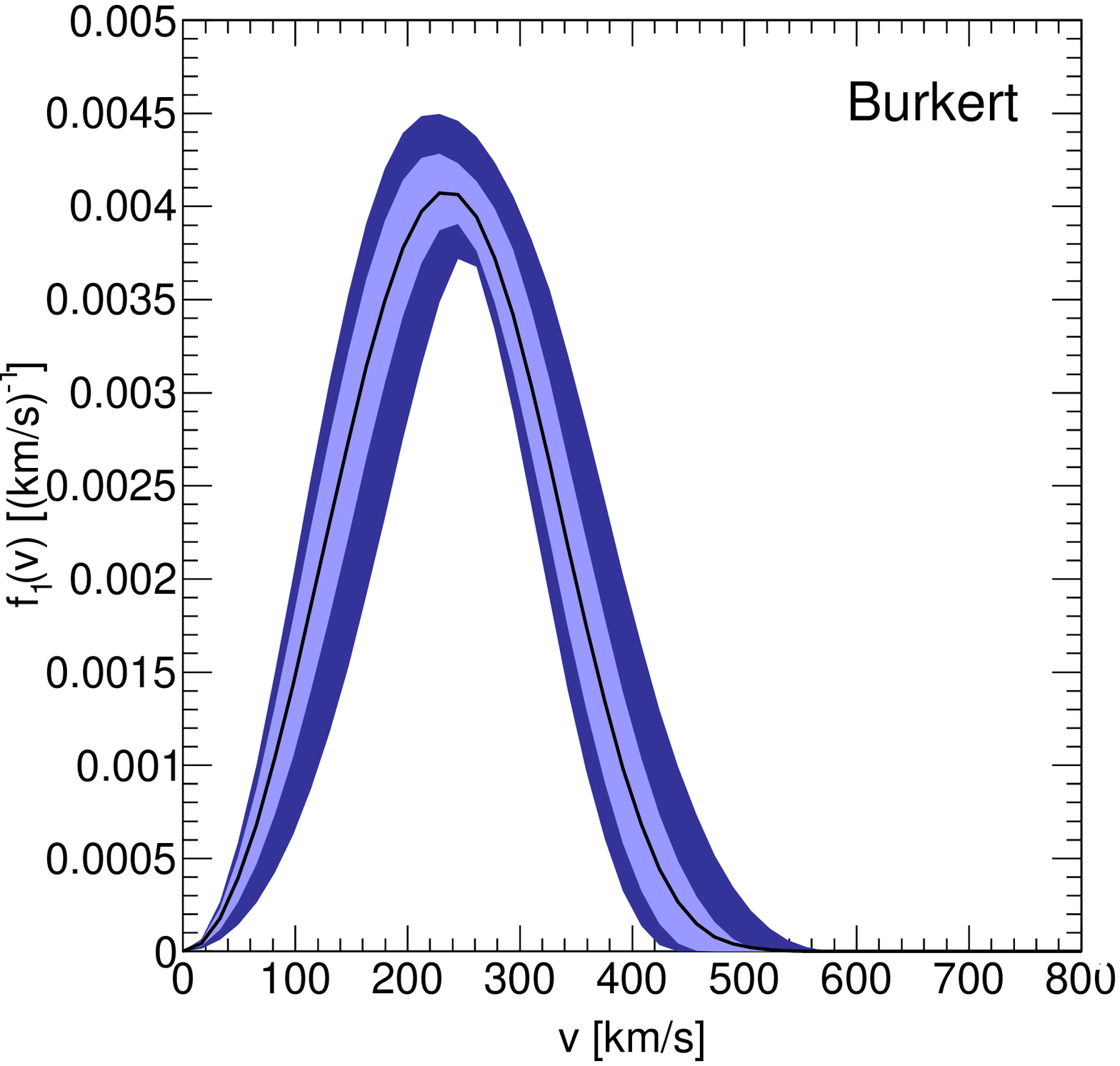}
\caption{\label{fig:Eddington} The probability distribution of the speed distribution, $f_1(v)$, resulting from the scan of the parameters of the MW mass model (see Sec. \ref{sec:results}) for, from left to right, the NFW, Einasto and Burkert DM halo profiles. In this case the speed distribution is obtained using the Eddington formalism, which assumes isotropy.  The light (dark) blue band indicates the 68\% (95\%) credible region, while the solid black line corresponds to the mean.}
\end{figure*}

\begin{figure*}
\includegraphics[width=0.32\textwidth]{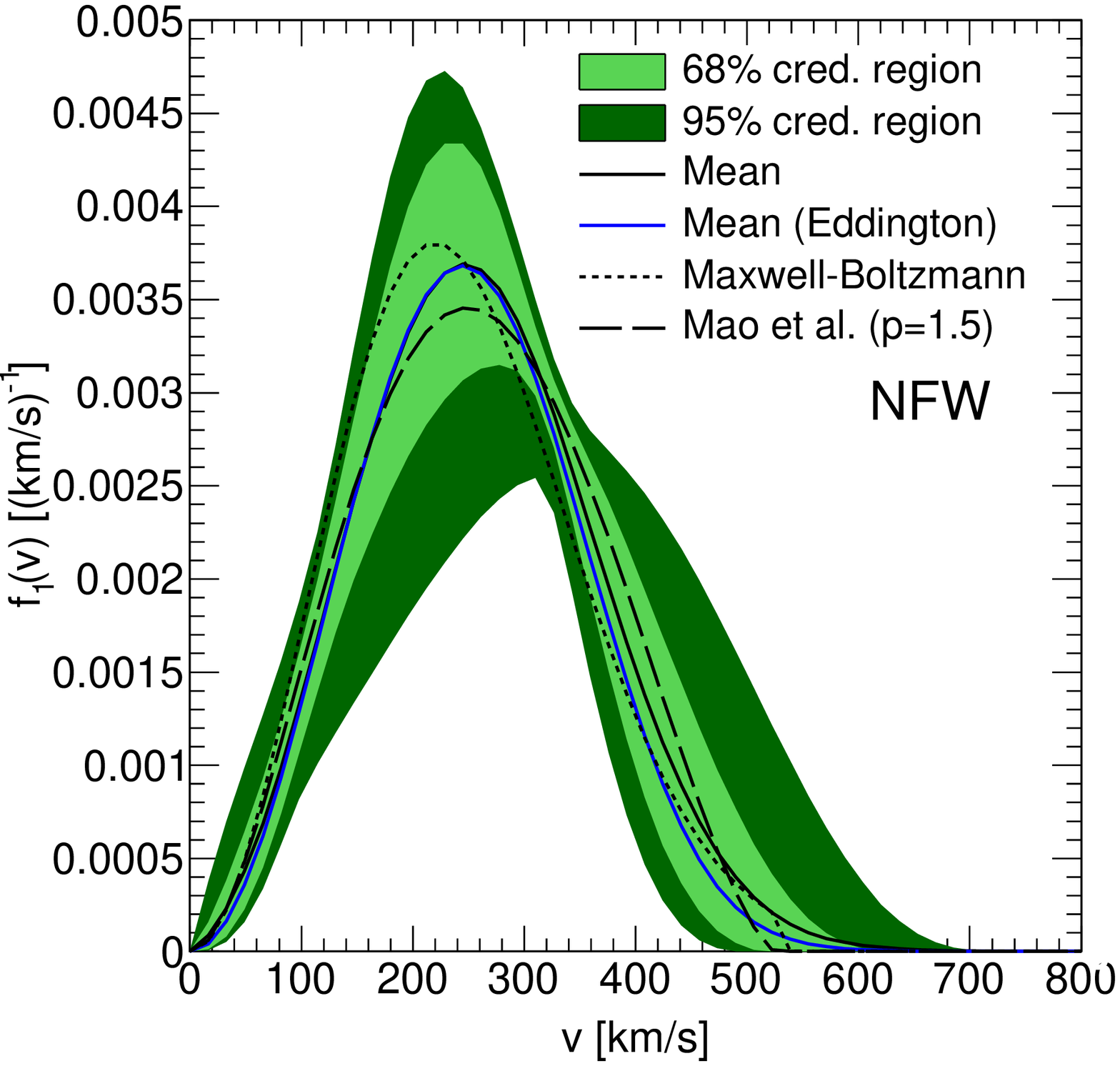}
\includegraphics[width=0.32\textwidth]{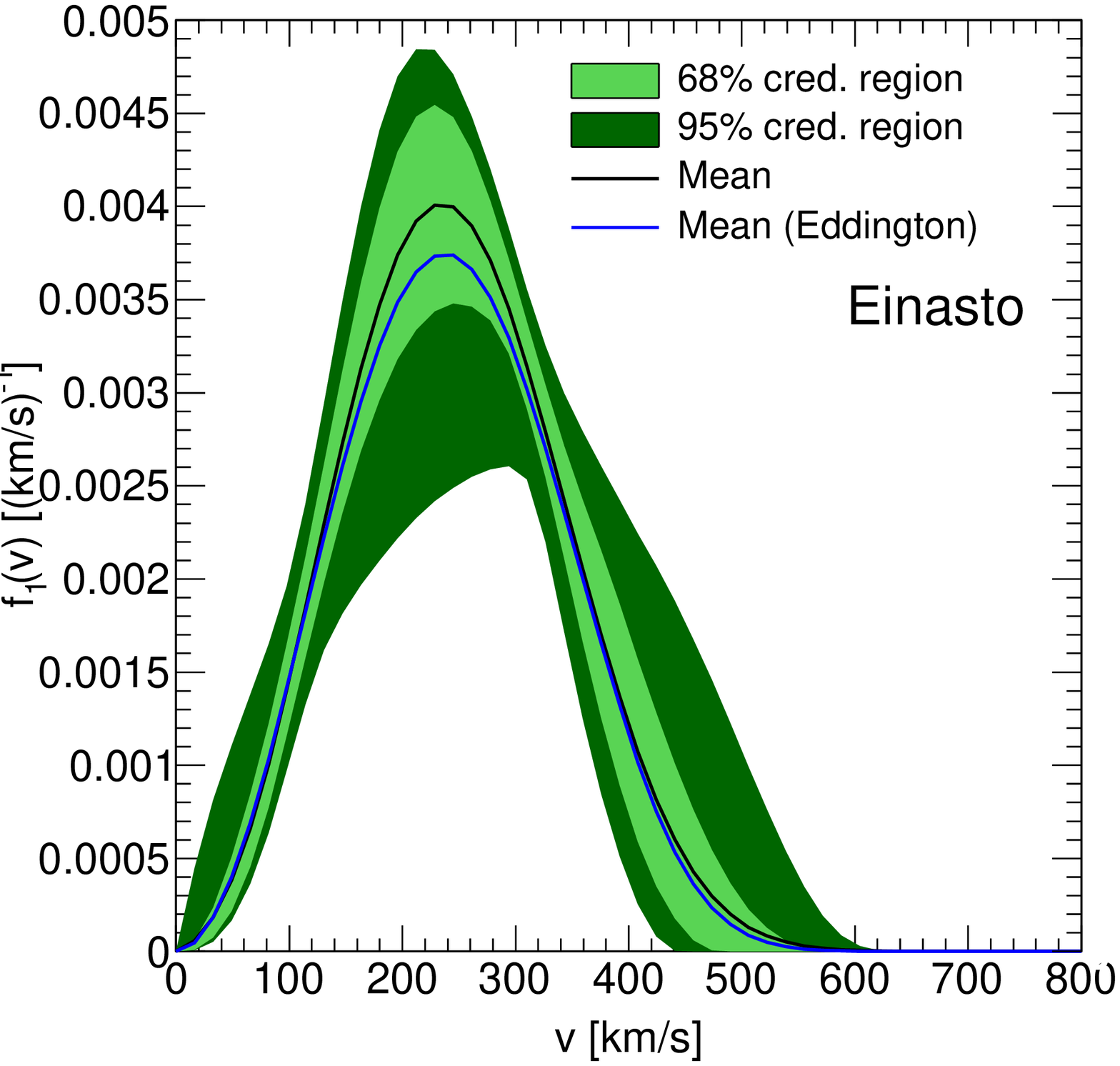}
\includegraphics[width=0.32\textwidth]{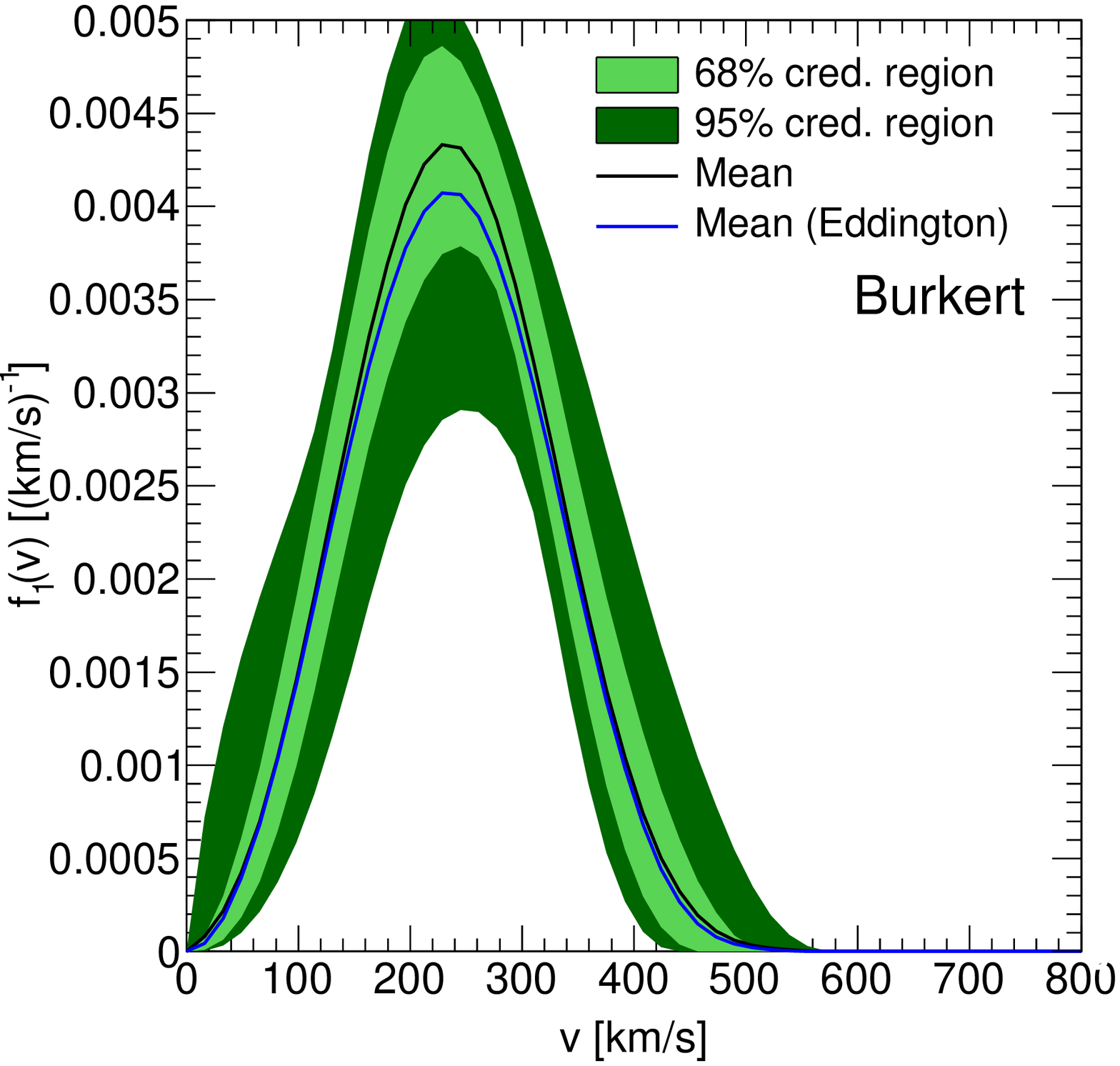}
\caption{\label{fig:ExtEddington}  As Fig. \ref{fig:Eddington} but for the speed distribution, $f_{1}(v)$, obtained using the procedure described in Sec. \ref{sec:velocity_distribution} which allows the DM halo to be anisotropic. The light (dark) green band indicates the 68\% (95\%) credible region, while the solid black line corresponds to the mean. The solid blue line indicates the mean of the distribution of $f_{1}(v)$ obtained from the Eddington procedure (as in Fig. \ref{fig:Eddington}). In the left panel (for the NFW DM halo), the dotted line is the Standard Halo Model Maxwell-Boltzmann distribution, Eq.~(\ref{shm}), while the dashed line shows the Mao et al. parametrization from Ref. \citep{Mao:2012hf}. In both cases the parameters are set to their mean values from Tab. \ref{tab:results_NFW}. }
\end{figure*}

In this section we present our results for the probability distribution of the 
speed distribution $f_1(v)$, derived from the scans discussed in 
Sec. \ref{sec:results}. In Fig. \ref{fig:Eddington}, $f_1(v)$ is obtained
through the Eddington formalism, assuming isotropy. The light (dark) 
blue bands show the 68\% (95\%) credible interval and the solid black line 
corresponds to the mean. The distribution is reconstructed with an
uncertainty of a factor of 2 (at 68\% credible interval) for speeds 
smaller than 500 ${\rm km \, s}^{-1}$. This is consistent with
the results of Refs. \citep{Catena:2011kv,Bhattacharjee:2012xm}. There seems 
to be a characteristic speed (around 300 ${\rm km \, s}^{-1}$) for which the 
speed distribution has a minimum uncertainty. This is particularly evident 
for the case of a NFW halo. We believe this is just a consequence of imposing 
the normalization of $f_1(v)$ to 1. Different mass models in the scan 
correspond to different gravitational potentials and, thus, to more peaked or 
extended speed distributions (see Fig. \ref{fig:fv}). However, increasing the 
amplitude of the peak has to lead to a less populated high-speed tail and, 
therefore, to a transition speed where smaller changes are 
observed\footnote{Note that, even if each $f_1(v)$ (corresponding to a 
specific mass models) has to be normalized to 1, the mean in 
Fig. \ref{fig:Eddington} (and in Fig. \ref{fig:ExtEddington}) does not 
necessarily integrate to 1, since it does not correspond to a specific mass 
model.}.

Our main result is displayed in Fig. \ref{fig:ExtEddington} where we show
the probability distribution of $f_1(v)$ obtained through the procedure 
summarized in the previous section (i.e. introducing a three-parameter 
form for $F_L(L)$, deriving $F_E(E)$ in a self-consistent way and marginalizing 
over the three parameters). As in Fig. \ref{fig:Eddington}, the light (dark) 
green shows the uncertainty in the reconstruction by means of the 68\% (95\%) 
credible interval and the solid black line corresponds to the mean. As 
expected, the reconstruction is worse than the isotropic case in 
Fig. \ref{fig:Eddington}, due to the presence of the three parameters in 
$F_L(L)$, however the speed distribution can still be determined with an 
accuracy of a factor of 4-5 for velocities smaller than 
500 ${\rm km \, s}^{-1}$. 
The most uncertain part is, as before, the high-speed tail. The region of 
small uncertainty at around 300 ${\rm km \, s}^{-1}$ is more evident than in 
the isotropic case and an additional region with small uncertainty 
appears at around 100 ${\rm km \, s}^{-1}$. These regions could already 
partially be seen in Fig. \ref{fig:fv}. The region at smaller (larger) 
speeds results from the effect of changing $\beta_0$ 
($\beta_\infty$)\footnote{Following the discussion given before, changing $L_0$ 
has similar effects to changing $\beta_0$ and $\beta_\infty$.}.

\begin{figure}
\includegraphics[width=0.45\textwidth]{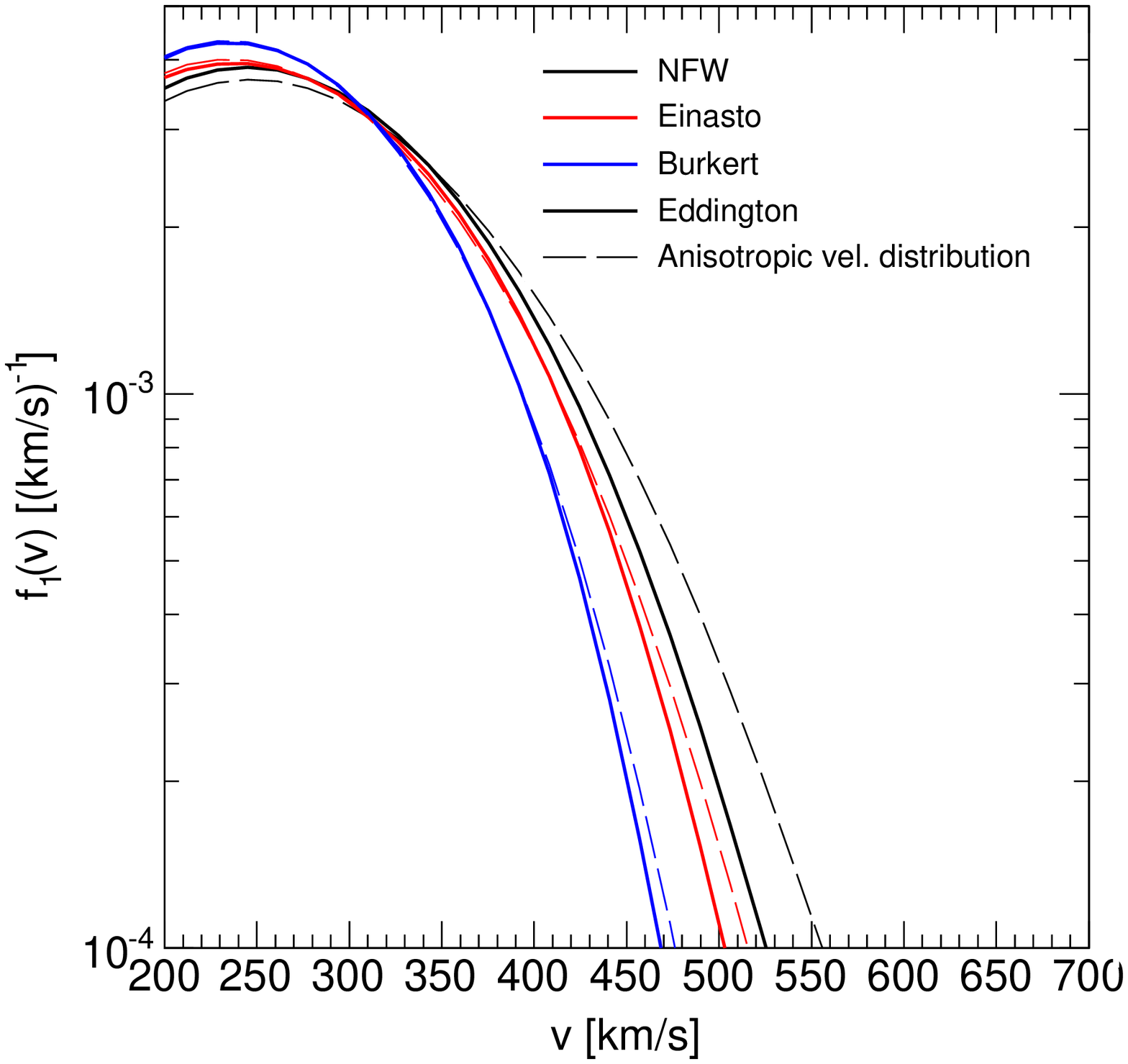}
\caption{\label{fig:fv_tail} The mean velocity distribution at large velocities for the NFW (black lines), Einasto (red lines) and Burkert (blue lines) DM profiles. Solid lines are for the isotropic case, calculated using the Eddington formalism, and dashed lines allowing anisotropy.}
\end{figure}

The solid blue lines show the speed distribution obtained through the 
Eddington formalism as in Fig. \ref{fig:Eddington}. For a more physical 
comparison the two distributions should be normalized to the same value, and 
once that is done the difference is small. For all of the DM halo profiles 
considered the tail of the speed distribution, beyond 
$\sim 500 \, {\rm km \, s}^{-1}$, is larger in the anisotropic case, and 
consequently the peak of the distribution is lower. The effect is 
approximately a factor of 2 for the NFW halo, while it is less pronounced for 
the Einasto and Burkert halos (see Fig. \ref{fig:fv_tail}). This is probably 
due to the marginalization over $\beta_\infty$ and the uncertainty in the 
behaviour of the DM halo at large radii. 

In the left panel of Fig. \ref{fig:ExtEddington}, for the case of the NFW 
halo, we also compare our results to the velocity distribution of the SHM 
(black dotted line) and the parametrization proposed by Mao et al. in
Refs. \citep{Mao:2012hf,Mao:2013nda} as a fit to the results of $N$-body 
simulations. 
This parameterization also provides a good fit to the DM speed distribution 
found in a hydrodynamical simulation of a MW-like galaxy containing baryons 
\citep{Kuhlen:2013tra}. In each case the circular and escape speeds are set 
to the mean values from Tab. \ref{tab:results_NFW} and we set $p$, which 
parameterizes the shape of the high speed cut-off, to $1.5$ as suggested by 
\citep{Mao:2013nda}.
As already pointed out in Ref. \citep{Mao:2012hf}, the self-consistent 
isotropic $f_1(v)$ has a lower high-velocity tail (solid blue line) than the 
Mao et al. phenomenological parameterization  (dashed black line in the upper 
left panel of Fig. \ref{fig:ExtEddington}). Our self-consistent anisotropic 
distribution (solid black) is however close to the Mao et al. 
parameterization (dashed black), which lies inside our 68\% credible band.

\section{Summary and conclusions}
\label{sec:conclusions}
In this paper we developed a mass model for the MW, with the goal of studying
the local DM velocity distribution in the case of a DM halo with an 
anisotropic velocity tensor.

Our MW mass model, inspired by Refs. \citep{Catena:2009mf,McMillan:2011wd}, 
assumes that the matter density of our own Galaxy can be written as a sum of 
four components: a disk, a combination of bulge and bar, a gaseous component 
and a DM halo. The free parameters entering in the modelling of these 
components are constrained by imposing a large set of astronomical 
observations (including the measurement of the proper motion of Sgr A$^\ast$, 
the local surface density and terminal velocities, as well as information 
derived from the motion of masers and halo stars and from microlensing). 

Such observations constrain the gravitational potential of the MW from which 
it is possible to reconstruct the DM phase-space density $F(E)$ (via the
Eddington equation), under the assumption of an isotropic velocity tensor.
The solution obtained will be self-consistent, since it depends entirely on
the potential of the system.

We extended this approach to the case of an anisotropic velocity tensor 
following the procedure introduced in Ref. \citep{Wojtak:2008mg}, where 
the phase-space density $F(E,L)$ is separable in the two variables 
$F(E,L)=F_E(E)F_L(L)$. Ref. \citep{Wojtak:2008mg} parametrized the 
$L$-dependent component in terms of three quantities ($\beta_0$, $\beta_\infty$ 
and $L_0$) and showed that, once this is done, self-consistent solutions can 
be obtained for each given mass model simply by inverting a Volterra integral 
equation. The phase-space density obtained provides good fit to the radial 
dependence of the velocity anisotropy parameter measured in simulated DM 
halos surrounding galaxy clusters.

We apply the same procedure to the description of the MW DM halo, noting
that this strategy extends the Eddington formalism to anisotropic scenarios 
(see also Ref. \citep{Bozorgnia:2013pua}). It follows the same general approach 
employed when dealing with mass models of the MW: unknown 
quantities (e.g. the density profiles of the matter components of the MW) are parameterized and 
the parameters are constrained by a set of observations. There 
are no observations that directly constrain the three parameters introduced 
for $F_L(L)$, therefore they must be marginalized over. This could, in 
principle, spoil any hope to reconstruct the velocity distribution if $f_1(v)$ 
were to vary considerably as $\beta_0$, $\beta_\infty$ and $L_0$ are varied.

Our main conclusions are:
\begin{itemize}
\item While $F_E(E)$ depends strongly on the values chosen for $\beta_0$, 
$\beta_\infty$ and $L_0$, this is not the case for $f_1(v)$ and an acceptable 
reconstruction can still be achieved.
\item The precision reached is, as expected, worse than when the Eddington
formalism, which assumes isotropy, is used. However $f_1(v)$ is determined 
within a factor of a few (less than a factor of 5 below 500 
${\rm km \, s}^{-1}$), independently of which profile is chosen for the 
DM halo. The largest uncertainties are in the high-velocity tail, while there 
are two ``sweet spots'' (around 100 and 300 ${\rm km \, s}^{-1}$) where the 
precision is better than $10-15\%$. This is a consequence of the marginalization 
over $\beta_0$, $\beta_\infty$ and $L_0$ and the fact that $f_1(v)$ is, by 
definition, normalized to 1.
\item The mean value of the distribution for the reconstructed $f_1(v)$ is very
similar to that obtained assuming isotropy, apart from in the high speed tail. 
For high speeds the mean anisotropic speed distribution is larger (up to a factor of 2 for
of a NFW halo) than the mean isotropic distribution derived with the Eddington formula 
(consequently the peak in the distribution is also lower).
\item The parametrization of $f_1(v)$ introduced in 
Refs. \citep{Mao:2012hf,Mao:2013nda}, as a good fit to simulated DM halos, lies 
inside the band of our 68\% credible interval.
\end{itemize}

Uncertainties in the local velocity distribution propagate in the analysis of 
direct detection data and may translate into large uncertainties in the 
reconstruction of the mass and interaction cross section of the DM particle, 
with the possibility of mis-reconstructions and biases (see, e.g., 
Refs. \citep{Fairbairn:2008gz,Strigari:2009zb,Peter:2011eu,Kavanagh:2012nr,
Pato:2012fw,Kavanagh:2013wba,Peter:2013aha}, just to cite a few). Different 
experiments are sensitive to different regions of $f_1(v)$ so that they are 
affected differently by changes in the velocity distribution. Therefore 
realistic modelling of $f_1(v)$ is crucial when comparing results from 
different experiments.

The procedure presented here relaxes the assumption of isotropy in
the analysis of mass models of the MW and allows, at the same time, a complete
Bayesian assessment of the uncertainties involved and of the precision of the
reconstruction of $f_1(v)$ (see also Ref. \citep{Bozorgnia:2013pua}).
However, we have still assumed a spherical halo and a diagonal velocity tensor. 
$N$-body simulations and stellar dynamics have already demonstrated that 
these remaining assumptions are not true. The development of more general 
strategies, allowing a more flexible and realistic description of the MW, is 
therefore required. The present paper represents the first step towards this 
goal.

Also, realistic and theoretically-unbiased analysis techniques will be
required in the near future, given the amount of information that will be
delivered and the precision that will be reached by upcoming stellar surveys 
(e.g. the GAIA satellite \citep{Perryman:2001sp,Eyer:2013wba}) or by the 
new data releases of already operational surveys, e.g. SEGUE 
\citep{Yanny:2009kg}, RAVE \citep{Siebert:2011ej} and PanSTARRS 
\citep{Kaiser:2002zz}.

\section*{Acknowledgments}
It is a pleasure to thank R. Catena for his help in developing the mass model 
and his comments on the paper. We also want to thank P. McMillan and R. Wojtak 
for very helpful discussions. We gratefully acknowledge the support of the 
Leverhulme Trust and the access to the University of Nottingham High 
Performance Computing Facility. M. F. also acknowledges the support of the 
project MultiDark CSD2009-00064.

\bibliography{bibliography}

\end{document}